# Stress management in composite biopolymer networks

Federica Burla[1][†], Justin Tauber[2][†], Simone Dussi[2]\*, Jasper van der Gucht[2] & Gijsje H. Koenderink[1]\*

[1] AMOLF, Department of Living Matter, Biological Soft Matter group, Science Park 104, 1098 XG Amsterdam, the Netherlands

[2] Physical Chemistry and Soft Matter, Wageningen University & Research, Stippeneng 4, 6708 WE Wageningen, Netherlands.

[†] These authors contributed equally to this work.

\* Corresponding authors.

Email addresses: g.koenderink@amolf.nl, simone.dussi@wur.nl

**Living tissues show an extraordinary adaptiveness to strain, which is crucial for their proper biological functioning[1,2]. The physical origin of this mechanical behaviour has been widely investigated using reconstituted networks of collagen fibres, the principal load-bearing component of tissues[3,4,5]. However, collagen fibres in tissues are embedded in a soft hydrated polysaccharide matrix which generates substantial internal stresses whose effect on tissue mechanics is unknown[6,7,8]. Here, by combining mechanical measurements and computer**

**simulations, we show that networks composed of collagen fibres and a hyaluronan matrix exhibit synergistic mechanics characterized by an enhanced stiffness and delayed strain-stiffening. We demonstrate that the polysaccharide matrix has a dual effect on the composite response involving both internal stress and elastic reinforcement. Our findings elucidate how tissues can tune their strain-sensitivity over a wide range and provide a novel design principle for synthetic materials with programmable mechanical properties.**

The soft tissues in our body such as the skin, muscles and arteries have a striking ability to switch from being soft at small deformations to being stiff at high deformations[9]. This adaptive response to strain enables tissues to accommodate dynamic processes such as cell proliferation and migration[10], while preventing tissue rupture[11]. Recent biophysical studies have revealed that the main determinant of tissue strain-stiffening is collagen, a scaffolding protein that forms fibrillar networks[12]. Collagen networks have the intrinsic ability to undergo a transition from soft to rigid when strained[13] because at the coarse-grained level of fibres they have a sub-isostatic architecture, meaning that the average number of fibres meeting at each junction (3 for branches and 4 for crosslinked fibres) is below the Maxwell stability criterion of 6 for a network of springs[14]. Fibrous networks are soft at small strains because they respond primarily by fibre bending, but stiffen at large deformations as the fibres align along the principal direction of strain and start to stretch[4,5,15]. However, collagen networks within tissues are always embedded in a soft hydrated matrix comprised of polysaccharides and glycosylated proteins whose mechanical role is unknown[16]. Recent computational models suggest that these matrix polymers might elastically reinforce the fibrillar collagen matrix[17,18]. However, the role of the matrix is likely more complex, because the constituent molecules carry large negative charges[19]. It has been shown that the matrix consequently generates substantial mechanical stresses, especially in cartilaginous tissues[6,7,8], which could potentially have a large impact on tissue mechanics given the strong strain-sensitivity of collagen.

To understand the interplay between elastic effects and internal stress in the mechanical response of tissues, we reconstitute biomimetic composites from two paradigmatic biopolymers: collagen and hyaluronan (Figure 1a). Collagen type I is the most ubiquitous member of the collagen family[12], and hyaluronan is abundant in healthy tissues and upregulated in solid tumours[20]. Rheology measurements show that collagen and hyaluronan individually respond rather differently to an applied shear strain (Figure 1b). Collagen networks stiffen with strain above a threshold of about 5-10%, as expected from their subisostatic fibrous architecture. By contrast, crosslinked hyaluronan networks exhibit a mechanical response that to a good approximation is linear in the range of strains that we consider (Figure 1b). Unlike collagen, hyaluronan polymers are flexible, with a persistence length of only 5-10 nm[21]. Their elastic properties are therefore determined by the chain entropy and the network response is linear until the long chains are pulled taut.

Having established the individual networks response, we generate composites by polymerizing collagen monomers in the presence of a mixture of hyaluronan chains and PEGDA crosslinkers. Collagen polymerization and hyaluronan crosslinking occur concurrently, resulting in an interpenetrating double network. The final collagen network looks almost identical to its pure collagen counterpart (see Supplementary Figure 1) and hyaluronan forms a background which is uniform over length scales much smaller than the collagen network mesh size (see Supplementary Figure 2), suggesting that the two networks do not interact, as confirmed by biochemical assays (see Supplementary Figure 3).

To test the impact of hyaluronan on the strain-stiffening response of collagen, we perform rheology measurements on composites with a fixed 1 mg/mL collagen concentration and hyaluronan concentrations $c_{HA}$ ranging from 0 to 7 mg/mL. The composites strain-stiffen, but with a strong dependence on the hyaluronan content (Figure 1b). The addition of hyaluronan results in a striking enhancement of the linear elastic modulus $G_0$ beyond the sum of the moduli of the individual components (Figure 1c). This synergistic enhancement appears to be at a maximum around 4 mg/mL, beyond the concentration at which hyaluronan first forms a percolating network (see Supplementary compared to pure collagen (Figure 1d). Both the onset strain $\gamma_0$, where non-linearity first sets in, and the critical strain $\gamma_c$, where collagen fibres complete the transition from a bending- to stretching-dominated mode of deformation, shift upwards with increasing hyaluronan concentration (Figure 1d).

Figure 4). In the nonlinear regime, the composite networks have a strongly delayed stiffening response

In pure collagen networks, one important factor influencing the onset of strain-stiffening is the network connectivity[5,22]. To test whether hyaluronan impacts the connectivity of collagen networks, we perform delayed crosslinking experiments, adding PEGDA crosslinks only after collagen polymerization is complete. As the mechanics of the composite networks is very similar irrespective whether hyaluronan is crosslinked during or after collagen polymerization (see Supplementary Figure 5), we conclude that the delayed strain-stiffening of composites is not caused by changes in collagen network connectivity.

Another important determinant of the onset of strain-stiffening in collagen networks is mechanical stress[4,5,23,24]. To test whether hyaluronan causes any build-up of stresses during network formation, we measure the time-dependence of the normal stress $\sigma_N$ that the composite system exerts on the top rheometer plate as it assembles (Figure 2a,b). Initially (t=0), collagen monomers and hyaluronan chains form a viscous solution. As collagen starts to assemble in fibres (Figure 2a) and hyaluronan starts to crosslink (t=20 min), $G_0$ and $\sigma_N$ increase simultaneously (Figure 2b). The positive sign of the normal stress indicates swelling of the polymerizing gel. As hyaluronan continues crosslinking and collagen continues polymerizing (t=60 min), the normal force reverses and decreases, eventually reaching negative values, indicating gel contraction. A similar normal stress evolution is observed for hyaluronan alone, whereas collagen develops no observable normal stress (see Supplementary Figure 6). We interpret the normal force build-up as a competition between the electrostatic repulsion of the charged hyaluronan polymers and the intrinsic tendency of polymer chains to contract to maximize their configurational entropy. At the beginning of the crosslinking process, the electrostatic interactions dominate, causing the system to swell. As crosslinking progresses, the fraction of crosslinked hyaluronan chains increases and the associated contractile forces eventually take over. To investigate whether the contractile stresses that develop in the hyaluronan gel are transmitted to the collagen network, we performed localized laser ablation experiments, using a high intensity laser to disrupt the hyaluronan meshwork in the interstices of the collagen network. Localized removal of hyaluronan caused an immediate recoil of neighbouring collagen fibres, indicating the release of mechanical stress (Figure 2d-e and Supplementary Figure 7). Altogether, our observations seem to suggest that the hyaluronan matrix indeed exerts compressive stresses on the collagen fibres, as sketched in Figure 2f.

To test whether the internal stress generated by hyaluronan can explain the altered strain-stiffening response of composite networks, we use two-component network simulations. We model collagen as a sub-isostatic network of fibre segments of length $l_{0,1}$ with bending and stretching energy proportional to the constants $\kappa$ and $\mu_1$, respectively, while hyaluronan is modelled as a homogeneous mesh of springs of length $l_{0,2}$ and stretching energy proportional to $\mu_2$ (Figure 3a). To assess the effect of increasing hyaluronan concentration, we first vary the matrix stiffness $\mu_2$. We observe an increase of the linear modulus of the composite networks with increasing $\mu_2$, but no changes in the non-linear behaviour (see Supplementary Figure 10). Therefore, we include the contractile tendency of hyaluronan by reducing the rest length of the matrix springs $l_{0,2}$. Subsequently, we allow the entire system to establish a mechanical equilibrium between the contracting stresses in the hyaluronan matrix and the resisting bending forces of the collagen network by allowing the system to compress (Figure 3a and Supplementary Figures 8,9). Inducing compression by lowering $l_{0,2}$ while keeping the matrix stiffness $\mu_2$ fixed has little effect on the linear regime, but delays the onset of strain stiffening (see Supplementary Figure 10). When we simultaneously increase $\mu_2$ to account for elastic reinforcement and lower $l_{0,2}$ to account for internal stress build-up, we recover both the linear mechanical enhancement and the delay in strain-stiffening (Figure 3b and Supplementary Figures 11 and 12).

Our model suggests that hyaluronan tunes the mechanics of collagen networks through a combination of elastic reinforcement and internal prestress. This implies that the synergistic mechanics of the composites should be highly dependent on the fibre rigidity $\tilde{\kappa}$ (the ratio between $\kappa$ and $\mu_1$). Specifically, fibres more resistant to bending (larger $\tilde{\kappa}$) should be less sensitive to stresses arising from the background matrix. Furthermore, they should be less sensitive to the elastic reinforcement, given their already large bending rigidity. To test this prediction, we use a different type of collagen that forms stiffer fibres thanks to molecular crosslinking mediated by non-helical peptide extensions of the triple helix known as telopeptides[25]. By fitting the mechanical response curves to an analytical expression for fibrous networks[22], we indeed find the fibre rigidity for the telocollagen to be $\tilde{\kappa}=1\cdot10^{-4}$, larger than for the atelocollagen considered so far, where $\tilde{\kappa}=3\cdot10^{-5}$ (see Supplementary Figure 13). The strain-stiffening behaviour of networks formed by telocollagen is indeed less affected by hyaluronan, as compared to the atelocollagen, with smaller changes in the onset and critical strain, as well as smaller changes in the linear mechanical enhancement in the linear regime (Figure 3c, Supplementary Figures 14, 15). Therefore, the mechanical properties of the composite system exhibit a phase space governed by a balance between the bending rigidity of collagen and the dual effect of the hyaluronan matrix on linear elasticity and internal stress (Figure 3d and Supplementary Figure 16).

Finally, we test whether the mechanical response of the double networks can be mapped onto an effective single-component model, as recently proposed for the linear elastic regime[17]. Our experiments suggest that this mapping should extend to the nonlinear elastic regime, as we can rescale the strain-stiffening curves of composites onto the curve for pure collagen by normalizing the nonlinear modulus with $G_0$ and the shear stress with an effective stress $\sigma_{eff}$ (Figure 4a, see Supplementary Figure 17), which is manually computed and allows to overlap the curves. To capture the elastic reinforcement of collagen by the surrounding hyaluronan matrix in a single-component model, we assign an enhanced bending rigidity $\tilde{\kappa}_{eff}$ to the fibres, which accounts for the hindrance of fibre bending by the matrix[17]. Internal stresses are implemented by compressing the network and resetting the bending rest angles (Figure 4b and Supplementary Figures 18-19). This simple model allows us to quantitatively map all experimental data for both atelocollagen and telocollagen networks onto a mechanical phase space controlled by $\tilde{\kappa}_{eff}$ and the level of collagen compression (Figure 4c and Supplementary Figures 19-20). The mapping reveals that the compression is significantly higher for atelocollagen compared to telocollagen, consistent with the more elaborate double network model (Figure 3d). We furthermore find that the parameters controlling the linear and the non-linear regime are decoupled, with the linear modulus $G_0$ depending on $\tilde{\kappa}_{eff}$ and the critical strain $\gamma_c$ on the extent of compression.

We showed by experiments on tissue-mimetic composites that the combination of fibrous collagen networks with a soft polysaccharide hydrogel results in highly tuneable nonlinear mechanics. The soft gel tunes the mechanics by introducing both elastic reinforcement and internal stresses. The stress-induced stiffening we observe is reminiscent of the active stiffening of cytoskeletal and extracellular matrix networks by contractile molecular motors or cells[26,27,28], but with the distinction that here energy consumption is not required. The minimal computational model we developed can be applied also to other soft tissues such as the plant cell wall, where a rigid cellulose network is combined with a soft pectin matrix[29]. Our findings not only elucidate how biology combines biopolymers with complementary properties to finely-tune the mechanics of living tissues, but also provide a new avenue for the design of synthetic elastic materials. The bio-inspired concept of combining semiflexible and flexible polymer networks to improve mechanical properties has been recently translated into material science[30]. Here, we showed that internal stress generation introduces a powerful control knob to tune the mechanical response of a material. This principle is irrespective of the nature of the stress source, and can therefore also be implemented in synthetic materials by using pH- or temperature-responsive components.

**References**

1. Fung, Y.C. Elasticity of soft tissues in simple elongation. *Am. J. Physiol.* **213(6),** 1532-1544 (1967).

2. Hall, M.S. *et al.* Fibrous nonlinear elasticity enables positive mechanical feedback between cells and ECMs. *PNAS* **113**, 14043-14048 (2016).

3. Motte, S., Kaufman, L.J. Strain stiffening in collagen I networks. *Biopolymers* **99**, 35-46 (2013).

4. Licup, A.J. *et al.* Stress controls the mechanics of collagen networks. *PNAS* **112,** 9573-9578 (2015).

5. Sharma, A. *et al.* Strain-controlled criticality governs the nonlinear mechanics of fibre networks. *Nature Physics* **12,** 584–587 (2016).

6. Narmoneva, D.A., Wang, J.Y., Setton, L.A. Nonuniform swelling-induced residual strains in articular cartilage. *J Biomech*. **32,** 401-8 (1999).

7.  Michalek, A.J., Gardner-Morse, M.G., Iatridis, J.C. Large residual strains are present in the intervertebral disc annulus fibrosus in the unloaded state. *J Biomech.* **45,** 1227-31 (2012).

8.  Voutouri C, Stylianopoulos T. Accumulation of mechanical forces in tumors is related to hyaluronan content and tissue stiffness. *PLoS One* **13,** 0193801 (2018).

9.  Lanir, Y. Multi-scale structural modeling of soft tissues mechanics and mechanobiology , *J. Elast.* **129**, 7-48 (2017).

10. van Helvert, S., Storm, C., Friedl, P. Mechanoreciprocity in cell migration. *Nat. Cell Biol.* **20,** 8-20 (2018).

11. Storm, C., Pastore, J.J., MacKintosh, F.C., Lubensky, T.C., Janmey, P.A. Nonlinear elasticity in biological gels. *Nature* **435,** 191-194 (2005).

12. Holmes, D.F., Lu, Y., Starborg, T., Kadler, K.E.  Collagen Fibril Assembly and Function. *Curr. Top. in Dev. Biol.* **130**, 107-142 (2018).

13. Broedersz, C.P., and MacKintosh, F.C. Modeling semiflexible polymer networks. *Rev. Mod. Phys.* **86,** 995 (2014).

14. Maxwell, J.C. On the calculation of the equilibrium and stiffness of frames. *Phil. Mag.* **27**, 294-299 (1864).

15. Onck, P.R., Koeman, T., van Dillen, T., van der Giessen, E. Alternative Explanation of Stiffening in Cross-Linked Semiflexible Networks. *Phys. Rev. Lett.* **95**, 178102 (2005).

16. Mouw, J.K., Ou, G., Weaver, V.M. Extracellular matrix assembly: a multiscale deconstruction. *Nat. Rev. Mol. Cell Biol.* **15,** 771-785 (2014).

17. van Doorn, J.M., Lageschaar, L., Sprakel, J., van der Gucht, J. Criticality and mechanical enhancement in composite fiber networks. *Phys. Rev. E* **95,** 042503 (2017).

18. Shahsavari, A. S. & Picu, R. C. Exceptional stiffening in composite fibre networks. *Phys. Rev. E* **92**, 012401 (2015).

19. Lai,V.K. *et al.* Swelling of Collagen-Hyaluronic Acid Co-Gels: An In Vitro Residual Stress Model. *Ann Biomed Eng.* **44,** 2984-2993 (2016).

20. Rankin, K.S., Frankel, D. Hyaluronan in cancer - from the naked mole rat to nanoparticle therapy. *Soft Matter* **12,** 3841-8 (2016).

21. Berezney, J.P., Saleh, O.A. Electrostatic Effects on the Conformation and Elasticity of Hyaluronic Acid, a Moderately Flexible Polyelectrolyte. *Macromolecules* **50**, 1085-1089 (2017).

22. Jansen, K.A. *et al.* The Role of Network Architecture in Collagen Mechanics. *Biophys. J.* **114**, 2665-2678 (2018).

23. Vahabi, M. *et al.* Elasticity of fibrous networks under uniaxial prestress. *Soft Matter* **12,** 5050-5060 (2016).

24. van Oosten, A.S.S.G. et al. Uncoupling shear and uniaxial elastic moduli of semiflexible biopolymer networks: compression-softening and stretch-stiffening. *Sci. Rep.* **6**, 19270 (2016).

25. Shayegan, M., Altinda, T., Kiefl, E., Forde, N.R. Intact Telopeptides Enhance Interactions between Collagens. *Biophys. J.* **111,** 2404-2416 (2016).

26. Koenderink, G.H. *et al.* An active biopolymer network controlled by molecular motors. *PNAS* **106**, 15192-15197 (2009).

27. Jansen, K.A., Bacabac, R.G., Piechocka, I.K., Koenderink, G.H. Cells actively stiffen fibrin networks by generating contractile stress. *Biophys J.* **105**, 2240-51 (2013).

28. Mizuno, D., Tardin, C., Schmidt, C.F., Mackintosh, F.C. Nonequilibrium mechanics of active cytoskeletal networks. *Science* **315,** 370-3 (2007).

29. Cosgrove, D.J. Growth of the plant cell wall. *Nat. Rev. Mol. Cell Biol.* **6**, 850–861 (2005).

30. Jaspers, M. *et al.* Nonlinear mechanics of hybrid polymer networks that mimic the complex mechanical environment of cells. *Nature Comm.* **8**, 15478 (2017).

## Acknowledgements

We thank Fred MacKintosh (Rice University, Texas), Eddie Pelan (University of Birmingham) and Simeon Stoyanov (Unilever B.V., The Netherlands) for many useful discussions, Abhinav Sharma (Leibniz Institute for Polymer Research, Dresden) for the MatLab script for determining the bending rigidity of collagen fibres from rheology data, David Nedrelow (University of Minnesota) for suggestions regarding the sample preparation, Kota Miura (EMBL, Heidelberg, Germany) for the Temporal Colour Code ImageJ plugin and Bas Overvelde (AMOLF, Amsterdam) for a critical reading of the manuscript. The work of F.B. and G.H.K. is part of the Industrial Partnership Programme *Hybrid Soft Materials* that is carried out under an agreement between Unilever Research and Development B.V. and the Netherlands Organisation for Scientific Research (NWO). The work of J.T., S.D. and J.v.d.G. is part of the SOFTBREAK project funded by the European Research Council (ERC Consolidator Grant).

## Author contributions

F.B. and J.T. contributed equally to the work. F.B. and G.H.K designed the experiments. F.B. performed and analysed the experiments. J.T. and S.D. designed, performed and analysed the simulations under the supervision of J.v.d.G. All authors interpreted and discussed results and co-wrote the paper.

## Competing interests

The authors declare no competing interests.

## Additional Information

**Supplementary Information** is available for this paper at ...

**Correspondence and requests for materials** should be addressed to G.H.K. or S.D.

# Methods

## Sample preparation

The tissue-mimetic composite system was prepared by mixing bovine dermal collagen type I atelocollagen (PureCol from CellSystems, supplied at 3.2 mg/mL in 0.01 M HCl) or telocollagen (TeloCol, from CellSystems, supplied at 3.1 mg/mL in 0.01 M HCl) and thiol-modified hyaluronan supplied as a lyophilised powder (Glycosil, 2B Scientific, 240 kDa) on ice. The networks were co-assembled in phosphate buffered saline (PBS, Sigma Aldrich) with the pH set to pH 7.4 using 0.1 M NaOH (Sigma Aldrich). Finally, the PEGDA (Polyethylene glycol diacrylate) crosslinker (Extralink PEGDA, 2B Scientific, 3.4 kDa) in a concentration ratio of 1:4 with hyaluronan  (molar ratio 20:1) was added just before inserting the sample in the rheometer or microscopy sample cells, to prevent premature crosslinking of the hyaluronan networks. The experiments were performed at a fixed concentration of 1 mg/mL collagen and at hyaluronan concentrations ranging from 0 to 7 mg/mL.

## Hyaluronan labeling

Hyaluronan was fluorescently labelled by coupling fluorescein-labelled maleimide (BDP FL maleimide, from Lumiprobe) with the thiol groups on the hyaluronan chains. Briefly, we dissolved the hyaluronan to a concentration of 10 mg/mL. We then dissolved the dye in dimethylsulfoxide (DMSO, Sigma Aldrich) to a concentration of 30 mg/mL and added it to the hyaluronan solution in a 10-fold molar excess. The reaction was allowed to proceed overnight at 4°C. The sample was finally gel-filtrated with a desalting column (PD-10, GE healthcare) to remove the excess dye.

**Rheology measurements**

All the experiments were performed on an Anton Paar Physica MCR501 rheometer, equipped with a cone plate geometry having a diameter of 40 mm and cone angle of 1°. We verified that the measurements were independent of gap size by repeating the experiments with a parallel plate geometry at gaps of 200 and 400 μm (data not shown). The samples were allowed to polymerize for two hours at a temperature of 37°C, maintained by a Peltier plate, while monitoring the evolution of the linear shear moduli with a small amplitude oscillatory strain (0.5% strain, 0.5 Hz). The steady-state values of the linear viscoelastic moduli were calculated as an average over the last ten data points of the polymerization curve, and the average reported is representative of at least 3 independent measurements. After polymerization, the nonlinear elastic response was measured using a well-established prestress protocol[1]. Briefly, a constant stress σ was applied for 30 s, to probe for network creep, and then an oscillatory stress δσ was superposed with an amplitude of 1/10 of σ. The resulting differential strain δγ was then used to calculate the differential (or tangent) modulus K' = δσ/δγ.

**Determination of linear mechanical enhancement, onset and critical strain**

The linear mechanical enhancement was determined by dividing the elastic modulus $G_0$ measured at the end of polymerization of the composite network ($G_{meas}$) by the sum ($G_{sum}$) of the elastic moduli $G_0$ of the two individual components. The error bar shown is calculated from error propagation.

The rheology data in the nonlinear regime were evaluated using a custom-written Python routine.

Consistent with a definition that was previously introduced in the context of fibrillar network mechanics[2], for the onset strain determination we first considered the curve K'/σ vs σ. Subsequently, we performed a cubic spline interpolation and determined the minimum of the resulting curve (see Supplementary Figure 21a). The associated stress was characterized as the onset stress $\sigma_o$, and the corresponding strain was taken to be the onset strain $\gamma_o$. To determine the critical strain, we calculated the cubic spline derivative of the curve log K' as a function of log γ. The strain at which this function attained its maximum was taken to be the critical strain $\gamma_c$ (see Supplementary Figure 21b). All the characteristic strain values are shown as average with standard error of the mean of at least three measurements.

**Delayed gelation experiments**

Delayed gelation experiments were performed by polymerizing collagen in the presence of hyaluronan in the rheometer for 40 minutes, before adding the PEGDA crosslinker around the trap. The crosslinker was then allowed to diffuse into the gel for 300 minutes before performing prestress measurements. To speed up the diffusion process, measurements were performed using a smaller cone plate geometry compared to the bulk of the measurements, with a diameter of 20 mm. We verified that the shear moduli of hyaluronan gels with the crosslinker mixed in before polymerization or added after polymerization were comparable, indicating that the crosslinker was indeed able to enter the gels (see Supplementary Figure 5).

**Microscopy experiments and image analysis**

The composite networks were imaged with an inverted Eclipse Ti microscope (Nikon, Tokyo, Japan) using a 100× oil immersion objective with a numerical aperture N.A. 1.49 (Nikon). The collagen networks were imaged in confocal reflectance microscopy mode, while hyaluronan was imaged by confocal fluorescence detection, and both components were imaged using a 488 Argon laser for illumination (Melles Griot, Albuquerque, NM). Images were acquired 10 μm above the bottom surface, over a depth of 30 μm and with a step size of 0.2 μm, and are shown as maximum intensity projections obtained with ImageJ[3]. Before imaging, the networks were allowed to polymerize for at least two hours in a sample holder composed of two coverslips (Menzel™ Microscope Coverslips 24x60mm, #1, Thermo Scientific) separated by a silicon chamber (Grace Bio-Labs CultureWell™ chambered coverglass, Sigma Aldrich). The sample holder was subsequently placed in a petri dish with humid tissue and closed by parafilm, in order to ensure constant humidity and then placed in a warm room (37°C) before observation. Time-lapse movies of network formation were obtained by polymerizing the networks directly on the microscope in a similar geometry as previously described, with the help of a custom-built temperature-controlled chamber. To determine the mesh size of the networks, we applied an algorithm[4] implemented in a custom-written Python routine. Briefly, images were background-subtracted and binarized with an Otsu threshold algorithm. Afterwards, the distance between white and dark pixels was counted along the rows and columns. Histograms of the distance distribution were then fitted with an exponential probability distribution and the mean value (converted from pixels to μm) was taken to be the average mesh size.

**Particle tracking**

Particle tracking was performed using latex fluorescent beads with a diameter of 0.4 μm (Reagent microspheres, Green visible Fluorescent Carboxylate modified, Duke Scientific). The beads were surface-functionalized with pluronic (Pluronic® F-127, Sigma Aldrich) to avoid interaction with the biopolymers following a published procedure[5]. The beads were added to the samples prior to polymerization, and after two hours, networks were imaged with the same microscope and same chamber as described above, at a distance of at least 10 µm above the bottom glass coverslip (1 fps acquisition, 100 ms exposure time). The particles' displacements in the x-y plane were determined with subpixel accuracy by analysing the movies with a custom written Python code based on Trackpy[6].

**Spin-down assay**

To check whether we could detect any association between hyaluronan and collagen, we performed a spin-down assay. Briefly, we co-polymerized collagen and hyaluronan (without PEGDA crosslinks) in an Eppendorf tube at 37°C for two hours. Afterwards, the collagen network was precipitated by centrifugation for 90 minutes at 8000g and the supernatant was collected. The presence of hyaluronan in the supernatant was assessed by measuring an UV-VIS absorption spectrum with a spectrophotometer (Thermo Scientific, Nanodrop 2000) and comparing this with the spectra for supernatants obtained from pure collagen and pure hyaluronan after a similar spin-down procedure.

**Determination of collagen bending rigidity**

To determine the bending rigidity of collagen fibres prepared from either atelocollagen or telocollagen, we measured the nonlinear elastic response of pure collagen networks prepared at 1 mg/mL using the prestress protocol described above. The dimensionless bending rigidity was extracted from the stress-stiffening curves by comparing them to curves obtained with the following analytical expression valid for submarginal fibrous networks described by Sharma *et al.*[7]:

$$\frac{\tilde{\kappa}}{|\Delta\gamma|^{\Phi}} \sim \frac{K'}{|\Delta\gamma|^f}\left(\pm 1 + \frac{K'^{\frac{1}{f}}}{|\Delta\gamma|}\right)^{(\Phi - f)} \quad \#(1)$$

where $\tilde{\kappa}$ is the free parameter corresponding to the dimensionless bending rigidity, K' is the measured value of differential modulus, $|\Delta\gamma|$ the distance between the considered $\gamma$ and the critical strain, and f and $\Phi$ critical exponents set by the network architecture, here taken to be 0.77 and 2.2 by comparing the measured values of onset and critical strain with published results[2].

**Laser ablation assay**

Laser ablation experiments were performed on a Nikon Ti Eclipse inverted microscope. The collagen network was imaged with confocal reflectance microscopy with the previously described setup. Next, the hyaluronan meshwork in the vicinity of a collagen fibre was ablated by a 30 second illumination with a circular spot of roughly 1-5 μm diameter using a pulsed infrared laser (Mai Tai, DeepSee, SpectraPhysics) operated at a wavelength of $\lambda = 700$ nm and a nominal power of 1400 mW. The collagen network was subsequently imaged again to check for recoil of collagen fibres in the vicinity of the ablation spot. Images are shown using an ImageJ hyperstack projection, such that fibres that remain immobile show up in white.

**Composite network simulations**

We model the tissue-like composite networks as an athermal and submarginal fibre network that represents the collagen network, embedded in a homogeneous (not diluted) matrix of linear elastic springs that represents the linearly elastic hyaluronan matrix[8,9] (see Supplementary Information). All simulations are performed on a lattice of L by L nodes, where L=50, with initial lattice spacing d=1.

For selected cases, we perform simulations with L=100 to confirm that size effects are within the statistical errors presented in the main text. The fibre network is modelled on a triangular lattice, using a dilution procedure to achieve a maximum local connectivity z of 4 bonds per node. As shown previously with similar models[2,10], it is possible to quantitatively describe the nonlinear elastic properties of reconstituted collagen networks, provided that model-dependent parameters, such as connectivity and number of crosslinks per fibre, are properly matched[2,10,11]. For these simulations we consider networks with an average connectivity $\langle z \rangle$=3.4. The total energy of the system is:

$$E = E_{stretch}^{fiber} + E_{bend}^{fiber} + E_{stretch}^{matrix} = \sum_{\langle ij \rangle} \frac{1}{2} \frac{\mu_1}{l_{0,1}} (l - l_{0,1})^2 + \sum_{\langle ijk \rangle} \frac{1}{2} \frac{\kappa_1}{l_{0,1}} (\theta - \theta_{0,1})^2 + \sum_{\langle ij \rangle} \frac{1}{2} \frac{\mu_2}{l_{0,2}} (l - l_{0,2})^2 \ \#(2)$$

where $\mu_1$ and $\kappa_1$ are the stretching and bending constant for the fibre segments, respectively, and $\mu_2$ is the stretching constant for the matrix segments. Nearest-neighbour nodes are indicated with indices $\langle ij \rangle$, and $l$ is the distance between the two nodes (spring length). Bending contributions are associated only to segments belonging to the same straight fibre, that are indicated with the triplet $\langle ijk \rangle$, $\theta$ denotes the angle between the triplet and $\theta_{0,1}$ the rest angle. The rest lengths $l_{0,1}$ and $l_{0,2}$ of the fibre and matrix segments are expressed in terms of the initial lattice spacing d (we use $l_{0,1}$ =d in all simulations), and we use the reduced quantities $\tilde{\mu}_2 = \mu_2/\mu_1$ and $\tilde{\kappa} = \kappa d^2/\mu_1$. We vary the matrix properties $\tilde{\mu}_2$ from $1 \cdot 10^{-5}$ to $1 \cdot 10^{-3}$ (see Supplementary Figure 10) while the fibre bending rigidity $\tilde{\kappa}$ is either $1 \cdot 10^{-4}$ or $3 \cdot 10^{-5}$, corresponding to the measured values for atelocollagen and telocollagen networks.

The mechanical behaviour of the composite networks is obtained in three steps. First, to simulate the compressive stress exerted by the hyaluronan matrix, we reduce $l_{0,2}$, which we vary from 0.50 to 1.00. Secondly, we allow the system to achieve an internal balance between the fibre network and the matrix via (isotropic) bulk compression in steps of 0.1% strain, with each step followed by energy minimization, until a minimum in the total energy as a function of bulk strain is reached. Thirdly, we perform a quasistatic shear simulation in steps of 0.1% strain, using Lees-Edwards boundary conditions in the y-direction and standard periodic boundary conditions in the x-direction. The energy minimization is performed using the FIRE algorithm[12]. The stress is obtained as $\sigma = \frac{1}{A}\frac{dE}{d\gamma}$ and the elastic shear modulus as $K' = \frac{1}{A}\frac{d^2E}{d\gamma^2}$, where A is the area of the system. We indicate the linear modulus (calculated for strains $\gamma < 2\%$) with $G_0$. We obtain the onset strain $\gamma_o$ and critical strain $\gamma_c$ that characterize the strain-stiffening response from the dependence of $\frac{K'}{\sigma}$ on $\sigma$. The first minimum corresponds to $(\gamma_o, \sigma_o)$, where $\sigma_o$ is the onset stress where stiffening sets in. The maximum corresponds to $(\gamma_c, \sigma_c)$, where $\sigma_c$ is the critical stress.

**Simulations of an effective single-component network**

To achieve a quantitative comparison with the experimental system, we develop an effective single-component model. In this case, we use a phantom fibre network with an average connectivity $\langle z \rangle = 3.2$, which for this network topology matches more closely the mechanical properties of the bare collagen system[2,7]. The presence of the hyaluronan matrix is implicitly implemented as follows. The effect of the matrix stiffness is captured by an increase of the fibre bending stiffness $\tilde{\kappa}$ to $\tilde{\kappa}_{eff}$, similar to recent simulations of the linear elastic properties of two-component composites[9]. To account for the local compression of the collagen network by the hyaluronan matrix, we isotropically compress the system in steps of 0.1% strain, followed by energy minimization. Subsequently, all the rest angles $\theta_0$ of the fibres are re-set to their value after compression to ensure mechanical equilibrium, meaning that the system retains its shape even if the boundaries are removed (see Supplementary Information for more details). To obtain the network mechanical response, a quasistatic shear simulation is performed in the same manner as described above.

**References Methods section**

1) Broedersz, C.P. *et al.* Measurement of nonlinear rheology of cross-linked biopolymer gels. *Soft Matter* **6**, 4120-4127 (2010).

2) Jansen, K.A. *et al.* The Role of Network Architecture in Collagen Mechanics. *Biophys. J.* **114**, 2665-2678 (2018).

3) Schindelin, J. *et al.* Fiji: an open-source platform for biological-image analysis. *Nature methods* **9,** 676-682 (2012).

4) Kaufman, L.J. *et al.* Glioma Expansion in Collagen I Matrices: Analyzing Collagen Concentration-Dependent Growth and Motility Patterns. *Biophys J.* **89**, 635–650 (2005).

5) Kim, A.J., Manoharan, V.N., Crocker, J.C. Swelling-Based Method for Preparing Stable, Functionalized Polymer Colloids. *JACS* **127,** 1592-1593 (2005).

6) Allan, D., Caswell, T., Keim, N., van der Wel, C. trackpy: Trackpy v0.3.2. (2016). doi:10.5281/zenodo.60550

7) Sharma, A. *et al.* Strain-controlled criticality governs the nonlinear mechanics of fibre networks. *Nat. Phys.* **12,** 584–587 (2016).

8) Broedersz, C., Mao, X., Lubensky, T.C., Mackintosh, F.C. Criticality and isostaticity in fibre networks. *Nat. Phys.* **7**, 983–988 (2011).

9) van Doorn, J.M., Lageschaar, L., Sprakel, J., van der Gucht, J. Criticality and mechanical enhancement in composite fiber networks. *Phys. Rev. E* **95,** 042503 (2017).

10) Licup, A.J. *et al.* Stress controls the mechanics of collagen networks. *PNAS* **112,** 9573-9578

(2015).

11) Lindström, S.B., Vader, D.A., Kulachenko, A., Weitz, D.A. Biopolymer network geometries: Characterization, regeneration and elastic properties. *Phys. Rev. E* **82**, 051905 (2010).

12) Bitzek, E., Koskinen, P., Gähler, F., Moseler, M., Gumbsch, P. Structural Relaxation Made Simple. *Phys. Rev. Lett.* **97**, 170201 (2006).

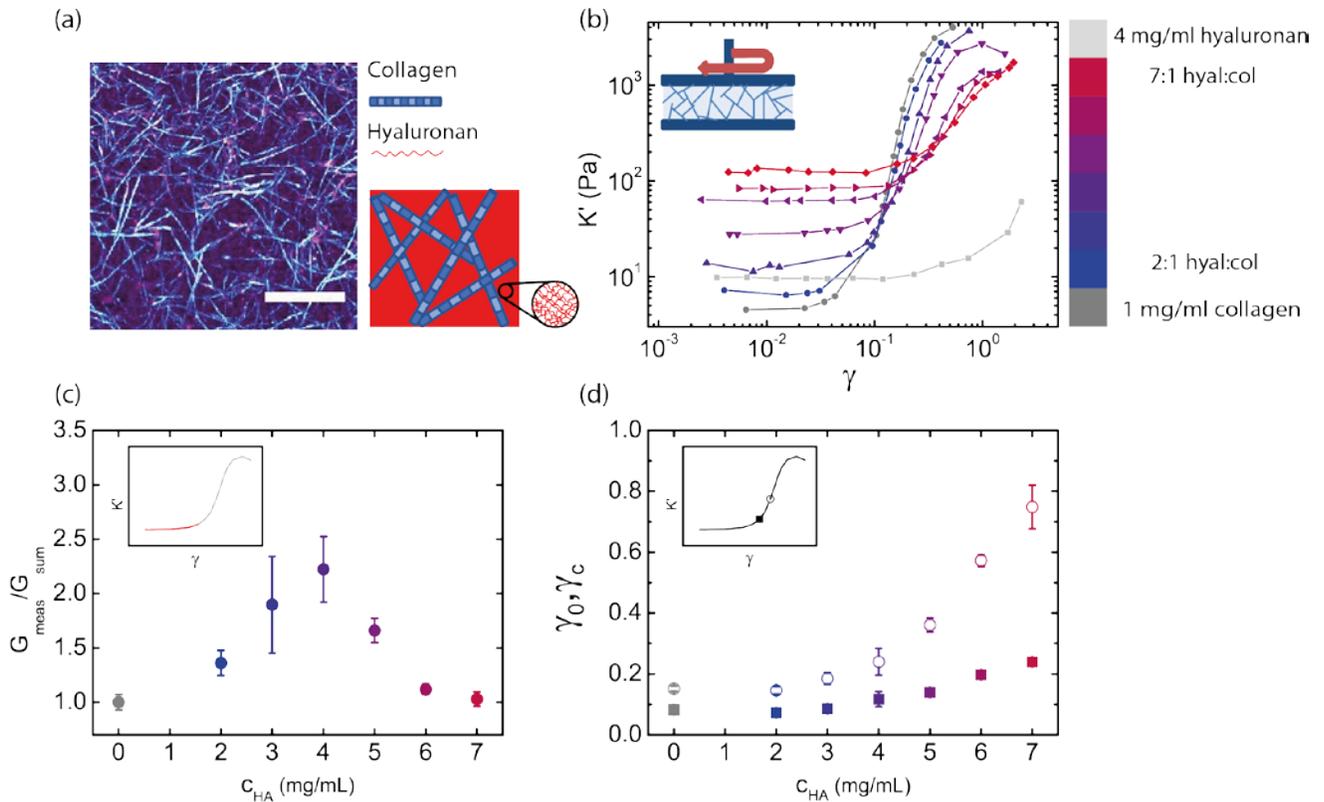

**Figure 1 | Composite collagen-hyaluronan networks as a minimal tissue-mimetic model system.**

(a) Confocal image together with a schematic of a composite network composed of a fibrous scaffold of collagen mixed with a soft hyaluronan gel that fills the interstices ($c_{col}$ = 1mg/mL, $c_{HA}$ = 4 mg/mL). (b) Differential elastic modulus, $K' = d\sigma/d\gamma$, as a function of applied strain for composites with a fixed concentration of collagen (1 mg/mL) and varying hyaluronan concentration (coloured curves). For comparison, pure collagen is shown in dark grey and hyaluronan (4 mg/mL) in light grey. Inset: the network is probed by shearing between two parallel plates. (c) Linear mechanical enhancement and (d) onset strain ($\gamma_0$, solid symbols) and critical strain ($\gamma_c$, open symbols) as a function of hyaluronan concentration, plotted with the same colour code as in (b). Data represent averages with standard error of the mean for at least three independent measurements.

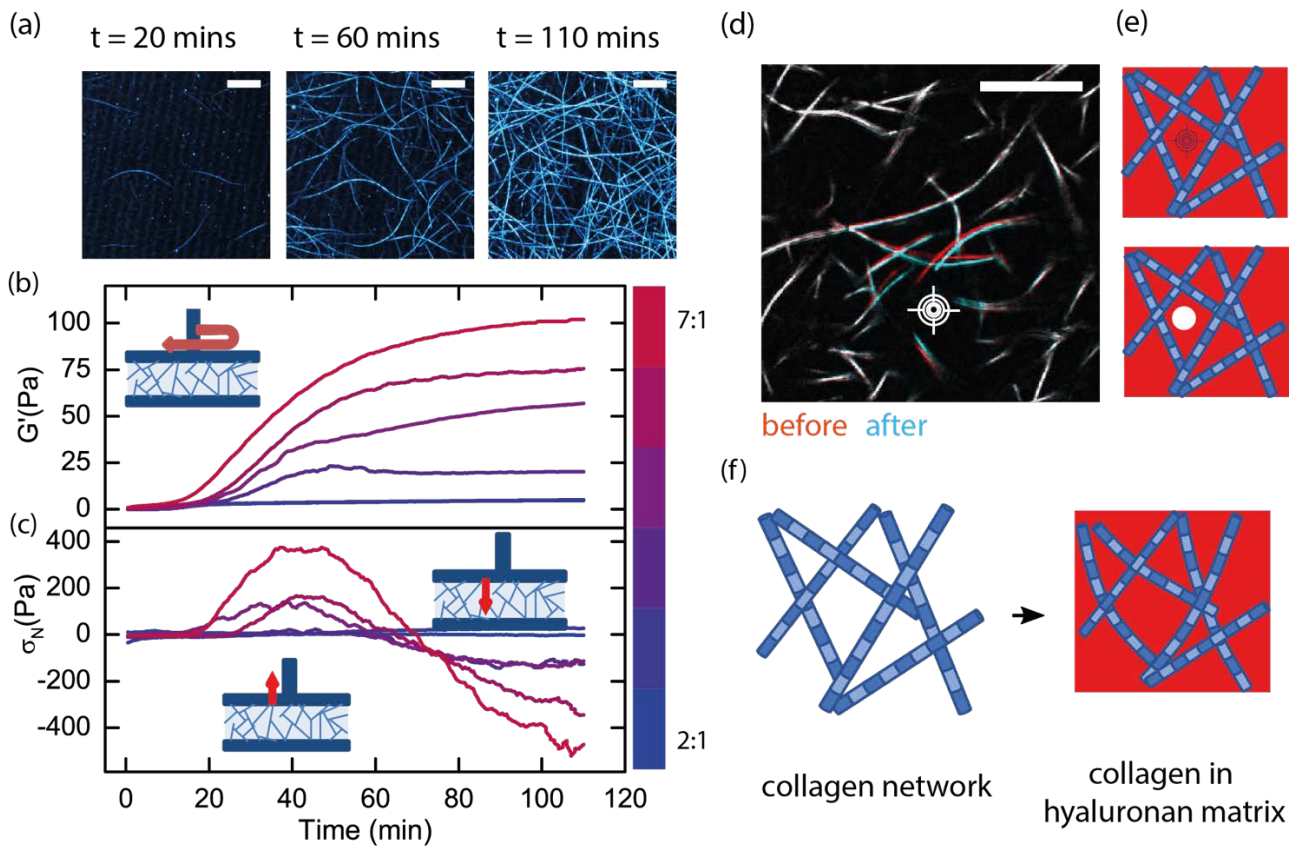

**Figure 2 | The hyaluronan gel prestresses the collagen network during gelation.**

(a) Z-projected confocal reflectance images of a composite network during the gelation process at three different time points, showing the formation of collagen fibres. Scale bars indicate 10 μm. (b) Evolution of the linear elastic modulus and (c) the corresponding normal force during network gelation for samples with a fixed collagen concentration (1 mg/mL) and different hyaluronan concentrations (from 2 mg/mL to 7 mg/mL). Note that the normal stress first increases, indicative of a swelling pressure, but afterwards decreases and reaches negative values, indicative of contractile stress. (d)

Superposed confocal images taken before (red) and after (blue) localized ablation of the hyaluronan matrix (target) with a strong laser, schematically shown in (e). The collagen fibres are everywhere immobile (visible as white) except in the vicinity of the laser spot, where they recoil. Scale bar is 5 μm. (f) Our data suggest that hyaluronan forms a contractile matrix that causes the collagen fibres to bend.

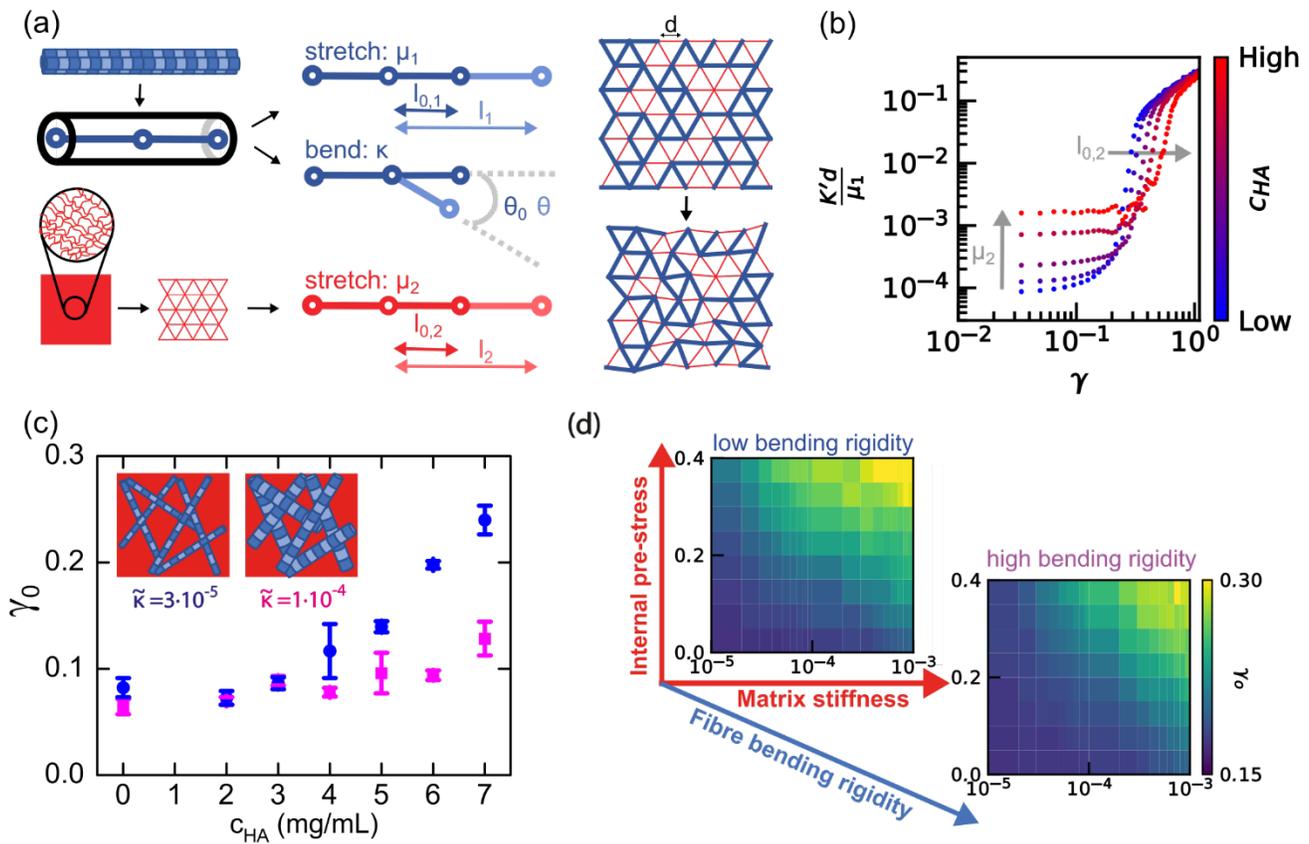

**Figure 3 | Computational modelling of two-component networks reveals that the composite mechanics depends on a balance between collagen bending and hyaluronan contraction.**

(a) The initial configuration is a sub-marginal fibre network (blue segments with stretching and bending resistance) embedded in a homogeneous flexible matrix (red segments, resistance to stretching only). To mimic the effect of hyaluronan contraction, we decrease the rest length $l_{0,2}$ of the matrix springs and reach mechanical equilibrium by allowing the network to compress. (b) Shear modulus as a function of strain for composite networks with varying matrix stiffness $\mu_2$ and varying internal pre-stress $l_{0,2}$. We qualitatively recover the experimentally observed dependence on hyaluronan concentration when we allow for a simultaneous increase in matrix stiffness and in compressive force. In this set of simulations, the average connectivity <z>=3.4 and the (rescaled) fibre bending stiffness $\tilde{\kappa}$ =3·10$^{-5}$ are kept fixed to the corresponding experimental values of the collagen network. (c) Experimentally measured onset strain for telocollagen (magenta squares) and atelocollagen (blue circles) networks as a function of hyaluronan concentration. (d) In simulations, $\gamma_0$ can be tuned by varying the collagen fibre rigidity and by independently varying the stiffness and internal stress of the hyaluronan gel, whereas experimentally the two matrix-related effects are inherently coupled when we vary hyaluronan concentration.

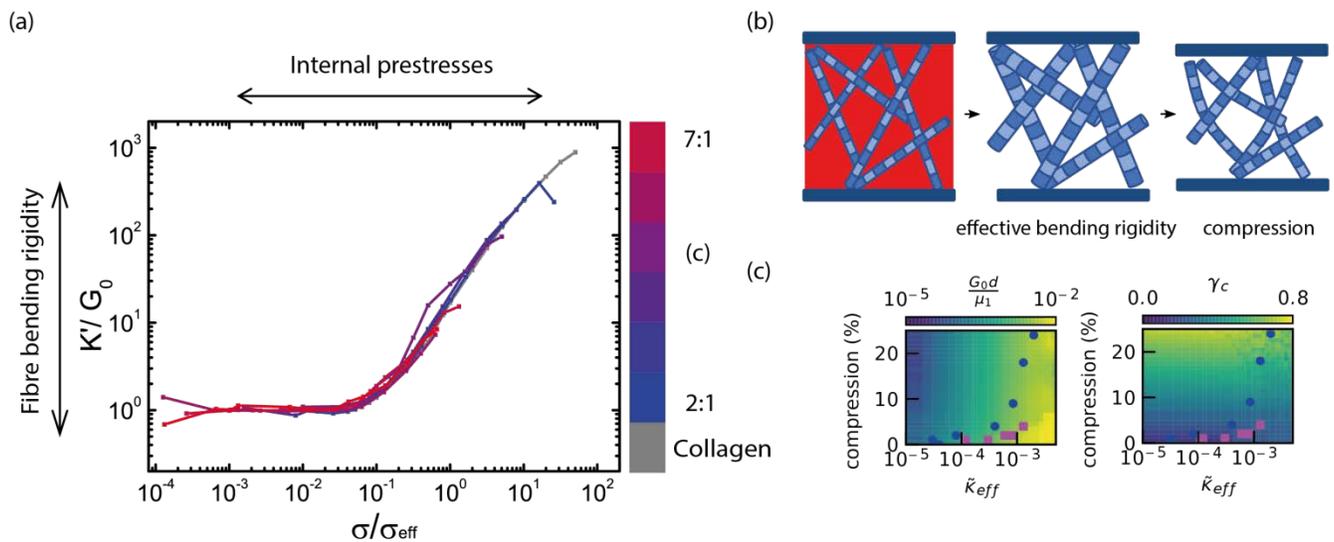

**Figure 4 | The mechanical response of composite networks can be mapped onto an effective single-component model.**

(a) The strain-stiffening curves of composite networks with varying hyaluronan concentration can be rescaled on top of the stiffening curve of a pure collagen network (grey) by normalizing the shear stress by an effective stress $\sigma_{eff}$ and the nonlinear modulus by $G_0$. $\sigma_{eff}$ is manually determined in order to be able to overlap the curves of the composites with the ones for collagen (see Supplementary Figure 17). The collagen concentration is fixed at 1 mg/mL. (b) The matrix effects can be incorporated in a single-component system by changing two properties of the collagen network: replacing the bending rigidity with an effective bending rigidity and compressing the network. (c) Map of $G_0$ and $\gamma_c$ as a function of $\tilde{\kappa}_{eff}$ and compression. The symbols (magenta squares for telocollagen and blue circles for atelocollagen) indicate the quantitative mapping of the experimental results on the single-component model.

# Supporting Information

# Stress management in composite biopolymer networks

Federica Burla[1][†], Justin Tauber[2][†], Simone Dussi*[2], Jasper van der Gucht[2] & Gijsje H. Koenderink*[1]

[1] AMOLF, Department of Living Matter, Biological Soft Matter group, Science Park 104, 1098 XG Amsterdam, the Netherlands

[2] Physical Chemistry and Soft Matter, Wageningen University & Research, Stippeneng 4, 6708 WE Wageningen, Netherlands.

[†] These authors contributed equally to this work.

* Corresponding authors. Email addresses: g.koenderink@amolf.nl, simone.dussi@wur.nl

# Simulations

## 1.1 Model assumptions

Several computational models have been proposed to describe collagen mechanics at different length scales.[1] We here adapt a spring network model based on the coarse-graining paradigm in which collagen fibres are treated as an athermal elastic material with a given Young's modulus and bending stiffness. Fibres are subsequently discretized in linearly elastic springs to which a stretching constant $\mu$ and bending constant $\kappa$ are associated. This minimal model has already been proven to correctly capture the physics behind the strain-stiffening of collagen networks.[2] Importantly, collagen fibre networks are sub-isostatic and therefore bending-dominated at small strains, so the strain-stiffening behaviour corresponds to the microscopic bending-to-stretching transition of the fibres. Our model could in principle be extended to capture sequential straightening and loading by modelling the fibres as springs with a non-linear stretching response. However, this non-linear response would be relevant only significantly beyond the critical strain at which the network mechanical response is stretching-dominated. To describe the hyaluronan network mechanics, we employ a uniform mesh and do not include bending contribution in the hyaluronan springs. This is justified by the fact that (i) hyaluronan is known to be a quite flexible polymer with a persistence length of only 5-10 nm; (ii) due to the large mismatch in network mesh-sizes, the much finer hyaluronan network appears as an homogenous matrix at the length-scales relevant for the mechanical response of collagen fibres; (iii) the experimental modulus versus strain curves of the pure hyaluronan network indicate that the hyaluronan network effectively behaves like a linearly elastic matrix in the range of strain interesting for the collagen and composite mechanics.

Details on the network geometries used in this work are given below. Examples of how the mesh geometry affects the strain-stiffening of sub-isostatic networks can be found in Refs. 3-6, where it is shown that this does not play a significant role. We use a lattice model because this is computationally much more efficient compared to e.g. Mikado models, especially in the case of two superimposed networks. Furthermore, the use of a triangular mesh rather than square or hexagonal allows us to tune the connectivity of the fibre network to physiological values between 3 and 4, which is not possible with a hexagonal lattice. The use of a square lattice also represents a limitation to study connectivities close to 4, since many system-spanning fibres would be present in the system giving rise to an unphysical mechanical response. Prior work on single-component systems showed that lattice models provide predictions that are fully consistent with 2D Mikado networks and 3D networks[2,3] Physically, this can be understood by the fact that the network mechanics is controlled only by a few parameters, such as the average connectivity $<z>$ and the average number of crosslinks per fibre $L_0/L_c$. Therefore, mechanically equivalent networks can be constructed using either a regular lattice or a disordered network.

## 1.2 Double network initialization

Initialization of double networks takes place in three sequential steps:

1) *Generation of the network geometry*: The collagen network is represented by a fibre network modelled on a triangular lattice composed of $L$ by $L$ nodes with lattice spacing $d$. We simulate on lattices of $L = 50$. A bond dilution procedure is designed to mimic the connectivity of collagen networks, where either two fibres meet at a node, resulting in a 3-fold branching point, or two fibres cross, resulting in a 4-fold crosslink.[7,8] Thus, the maximum connectivity at every node should be 4 as opposed to 6 for a regular undiluted triangular lattice. To fulfil this requirement, we use the following procedure (an adaptation of other recent simulations of collagen networks[9]): 1) we make a list of all the nodes with more than 4 connecting segments, 2) randomly choose a node in this list, 3) randomly remove one segment connecting to this node. We repeat this procedure until all nodes have a maximum of 4 connecting segments. Successively, we further dilute the network to reach the desired average connectivity ⟨z⟩ Here we perform simulations on networks with ⟨z⟩ = 3.4. All segments of the fibre network have a rest length equal to the lattice spacing $d$, and stretching or compression of these segments costs energy. In addition, all segments that are connected to each other in a straight line have a bending contribution. The hyaluronan matrix is instead represented by a homogeneous network of springs superposed on the same triangular lattice used to generate the submarginal fibre network. The fibre and matrix segments share the same nodes. Since the matrix segments are never diluted, every node has at least 6 matrix bonds connected to it and up to 4 fibre bonds. An example of a typical two-dimensional composite network obtained with this procedure is shown in the Supplementary Figure 9.

2) *Implementation of internal stresses before shear simulations*: To model the contractile stresses introduced by hyaluronan during the crosslinking process, we reduce the rest length $l_{0,2}$ of the segments of the homogeneous spring network, which initially equals $d$, in steps of $0.1d$. After every step, the energy of the complete system is minimized including both the fibre and the matrix segments.

3) *Mechanical equilibration before shear simulations*: Step 2 brings the system out of mechanical equilibrium. We define mechanical equilibrium as a situation in which the system would retain its shape if the boundaries were removed, or equivalently any small deformation would increase the total energy of the system. To re-establish mechanical equilibrium, we change the size of the simulation box. Specifically, we need to make the box smaller, to accommodate for the contractile tendency of the matrix. We therefore isotropically shrink the box in steps of 0.01% bulk strain. After every step the energy is minimized. As a result, the matrix relaxes, but the fibres bend in order to still fit in the reduced volume, as shown in Supplementary Figure 9. At small bulk strains, the increase in bending energy is outweighed by the decrease in matrix stretching energy. However, once the bulk strain reaches high enough values, the increase in bending energy dominates. These two opposing tendencies lead to a minimum in the total energy, as shown in Supplementary Figure 10. The corresponding compression is expressed as ($L_{init}$ – $L_{min}$)/$L_0$, where $L_{init}$ is the initial length of one of the axes of the simulation box and $L_{min}$ is the length once the energy minimum is reached. The amount of compression needed to achieve mechanical equilibrium depends on the stiffness and the reduction of the segment length in the matrix (Supplementary Figure 9) as well as on the fibre bending rigidity.

To obtain the mechanical response of the composite networks, we apply a shear strain up to 150% in steps of 0.1%. We use experimentally relevant values for the bending rigidity based on fits of a fibrous network model[7] to the strain-stiffening curves measured for atelocollagen ($\tilde{\kappa} = 3 \cdot 10^{-5}$) and telocollagen ($\tilde{\kappa} = 1 \cdot 10^{-4}$) networks. Since the experimental results show that the linear moduli of pure hyaluronan (Supplementary Figure 4) and collagen are of the same order of magnitude, we choose parameters in the simulations such that also here the linear moduli of the individual components are comparable (Supplementary Figure 10).

To calculate the modulus versus strain curves, the energy density versus strain curves are smoothened (running average with a range of 1% strain) and 100 equally spaced points are sampled via linear interpolation. The first derivative of this curve gives the stress, while the second derivative provides the modulus as a function of strain. From these data, the linear modulus $G_0$, onset strain for stiffening $\gamma_o$, and critical strain $\gamma_c$ are obtained for every individual curve. Data shown are averages over $N_{config} \sim 10$ samples.

### 1.3 Force equations used in the energy minimization algorithm

Energy minimization for our model is performed using the recently introduced FIRE algorithm that uses expressions for the force to find the minimum in the energy[10]. Here we present the equation for the stretching force and derive the equations for the different force components caused by bending the fibres. Nodes $i$ of the network have coordinates $\mathbf{x}_i = [x_{i,1}, x_{i,2}]$ and are connected by segments $\mathbf{r}_{ij} = \mathbf{x}_j - \mathbf{x}_i$. The force due to the stretching contribution acting upon node $\mathbf{x}_i$ along $\mathbf{r}_{ij}$ is defined as

$$f_{stretch, x_i} = -\frac{\mu_1}{l_{0,1}} \left( |\boldsymbol{r}_{ij}| - l_{0,1} \right) \#(1)$$

The expression for the bending force depends on a set of three nodes $\{\mathbf{x}_i, \mathbf{x}_j, \mathbf{x}_k\}$. For each node, the force equation is found via the gradient of the bending energy $-\nabla_{node} \mathrm{E}_{bend}$, where

$$E_{bend} = \frac{\kappa}{2l_{0,1}} (\Delta\theta)^2 = \frac{\kappa}{2l_{0,1}} (\theta - \theta_0)^2 \#(2)$$

To derive the force, we need an expression for $\theta$ as a function of the node coordinates $\{\mathbf{x}_i, \mathbf{x}_j, \mathbf{x}_k\}$. Here we use $\cos\theta = \frac{r_{ij} \cdot r_{jk}}{|r_{ij}||r_{jk}|}$. If $\theta = -\pi \dots 0$,

$$\theta_-(\boldsymbol{x}_i, \boldsymbol{x}_j, \boldsymbol{x}_k) = -\mathrm{acos}\frac{\boldsymbol{r}_{ij} \cdot \boldsymbol{r}_{jk}}{|\boldsymbol{r}_{ij}||\boldsymbol{r}_{jk}|} + \theta_{cor} \#(3)$$

and the force reads

$$f_{bend, x_{m,n}} = -\frac{\frac{\kappa * \left( \theta_-(\boldsymbol{x}_i, \boldsymbol{x}_j, \boldsymbol{x}_k) - \theta_0 \right)}{l_{0,ij} + l_{0,jk}} g_{x_{m,n}}(\boldsymbol{x}_i, \boldsymbol{x}_j, \boldsymbol{x}_k)}{\sqrt{1 - \left( -\frac{\boldsymbol{r}_{ij} \cdot \boldsymbol{r}_{jk}}{|\boldsymbol{r}_{ij}||\boldsymbol{r}_{jk}|} \right)^2}} \#(4)$$

or if $\theta = 0 \dots \pi$,

$$\theta_+(\boldsymbol{x}_i, \boldsymbol{x}_j, \boldsymbol{x}_k) = \mathrm{acos}\frac{\boldsymbol{r}_{ij} \cdot \boldsymbol{r}_{jk}}{|\boldsymbol{r}_{ij}||\boldsymbol{r}_{jk}|} - \theta_{cor} \#(5)$$

and the force reads

$$f_{bend,x_{m,n}} = \frac{\frac{\kappa * (\theta_+(\boldsymbol{x}_i, \boldsymbol{x}_j, \boldsymbol{x}_k) - \theta_0)}{l_{0,ij} + l_{0,jk}} g_{x_{m,n}}(\boldsymbol{x}_i, \boldsymbol{x}_j, \boldsymbol{x}_k)}{\sqrt{1 - \left(\frac{\boldsymbol{r}_{ij} \cdot \boldsymbol{r}_{jk}}{|\boldsymbol{r}_{ij}||\boldsymbol{r}_{jk}|}\right)^2}} \#(6)$$

Here, $\theta_{cor}$ is equal to $2\pi$ if $|\theta - \theta_0|$ is larger than $\pi$ and is 0 otherwise. The functions $g_{x_{m,n}}$ are specific for the node and the component of the force that is considered:

$$g_{x_{i,n}} = \frac{(x_{j,n} - x_{i,n})(\boldsymbol{r}_{ij} \cdot \boldsymbol{r}_{jk})}{|\boldsymbol{r}_{ij}|^3 |\boldsymbol{r}_{jk}|} + \frac{x_{j,n} - x_{k,n}}{|\boldsymbol{r}_{ij}||\boldsymbol{r}_{jk}|} \#(7)$$

$$g_{x_{j,n}} = \frac{(x_{i,n} - x_{j,n})(\boldsymbol{r}_{ji} \cdot \boldsymbol{r}_{jk})}{|\boldsymbol{r}_{ij}|^3 |\boldsymbol{r}_{jk}|} + \frac{x_{k,n} + x_{i,n} - 2x_{j,n}}{|\boldsymbol{r}_{ij}||\boldsymbol{r}_{jk}|} + \frac{(x_{k,n} - x_{j,n})(\boldsymbol{r}_{ji} \cdot \boldsymbol{r}_{jk})}{|\boldsymbol{r}_{ij}||\boldsymbol{r}_{jk}|^3} \#(8)$$

$$g_{x_{k,n}} = \frac{(x_{j,n} - x_{k,n})(\boldsymbol{r}_{ji} \cdot \boldsymbol{r}_{jk})}{|\boldsymbol{r}_{ij}||\boldsymbol{r}_{jk}|^3} + \frac{x_{j,n} - x_{i,n}}{|\boldsymbol{r}_{ij}||\boldsymbol{r}_{jk}|} \#(9)$$

## 1.4 Constructing an effective single-component model

Recent simulations[11] have shown that in the linear elastic regime, the effect of an elastic background network on the mechanics of a fibrous network can be mapped onto an effective single-component model. The basic idea is that the reinforcing effect of the matrix can be accounted for by a modified fibre bending rigidity, $\tilde{\kappa}_{eff} = \tilde{\kappa} + \tilde{\mu}_2 d^2$ for a triangular network, where $d$ is the lattice spacing. In this work, we tested this model also in the nonlinear regime on a phantom network geometry. As shown in Supplementary Figure 19, we find that an increase in $\tilde{\kappa}_{eff}$ only affects the linear regime.

To implement the second effect of a hyaluronan matrix on the collagen network, which is to exert a compressive stress, we therefore need to extend the effective single-component model. In the two-component model, the fibre network was compressed by the matrix as a result of the mechanical equilibrium obtained via energy minimization. In an effective single-component system, we can still perform a bulk compression to simulate the compressive effect of the matrix. This is done in steps of 0.1% bulk strain. Similar to the two-component simulations, we again observe bending of the fibres, as can be seen in Supplementary Figure 19. Furthermore, the introduction of compressive stress alone is able to capture the delay of non-linear stiffening that we observe experimentally in collagen/hyaluronan composites. We note that this effect of internal stress resembles the effect of external compression in fibre networks reported in literature[12]. However, our procedure does not yet correctly reproduce the linear regime behaviour. In fact, when we apply bulk compression to the whole system, fibres bend to account for the smaller system size, which results in a positive contribution to the energy. After the bulk compression we switch to a shear deformation. Because the system reorients along the shear direction, some of the bent fibres can straighten out again, leading to a lower system energy upon increasing the shear strain. This implies a negative shear stress and also a negative modulus. At higher shear strains, other fibres are forced to bend and the total energy increases again. The mechanical balance between the fibre network and the background matrix is therefore not yet correctly captured in this single-component model. To achieve an internal force balance, we reset all the fibre rest angles to the angles obtained after compression. In other words, the new rest state of the fibres of the single-component model is their bent configuration (Supplementary Figure 22). As bending is the only significant energy contribution after compression, this brings the energy of the system close to zero again. This procedure removes the negative shear stresses in the linear regime and still leads to a delay of stiffening in the nonlinear regime (Supplementary Figure 19).

## 1.5 Mapping the experimental data on the effective single-component model

The double network model reproduces the experimentally observed delay in stiffening of collagen networks upon the addition of a hyaluronan matrix, but only qualitatively. This is a consequence of a mismatch in geometry between the experimental system and the model. In particular, it can be explained by considering the average number of crosslinks per fibre, calculated as $L_0/l_c$, with $L_0$ the average fibre length and $l_c$ the average distance between crosslinks on a fibre. For the diluted triangular network, for connectivities between 3 and 4, we have $L_0/l_c \sim 2$, which is much lower than for geometries that have been used successfully in quantitative mapping of collagen elasticity[7,13]. As an example, Mikado networks have $L_0/l_c$ =11 at a connectivity of 3.6[13] and a triangular phantom lattice can have values of $L_0/l_c$ ranging from 2.59 to 10.99 for a connectivity between 3 and 4.[7] Unfortunately, the phantom lattice geometry is not convenient for explicit double-network simulations, since the phantomization procedure would lead to nodes with 12 matrix bonds and further assumptions on the springs connecting the phantom nodes would be necessary. Therefore, we performed double network simulations on diluted triangular network but effective single-component network simulations on phantom lattices.

We performed single-component network simulations for lattices with an average connectivity $\langle z \rangle$=3.2, chosen because the critical and onset strain at 0% compression ($\gamma_o$=0.068±0.007, $\gamma_c$=0.13±0.02) are closer to the strains measured for the collagen-hyaluronan composite ($c_{HA}$=2 mg/mL, $\gamma_o$=0.073±0.006, $\gamma_c$=0.15±0.01) compared to the results of our simulations obtained with $\langle z \rangle$=3.4 ($\gamma_o$=0.05, $\gamma_c$=0.09). Subsequently, we constructed maps of $G_0$, $\gamma_o$ and $\gamma_c$ as a function of bulk compression (26 points ranging from 0% to 25%) and effective bending rigidity (51 points ranging from $\tilde{\kappa}_{eff} = 1 \cdot 10^{-5} \dots 5 \cdot 10^{-3}$). We then mapped experimentally determined values of $G_0$, $\gamma_o$ and $\gamma_c$ obtained for composite networks with different HA concentrations onto the simulation results (Supplementary Figure 19). The goal of this quantitative mapping procedure is to find the relation between $c_{HA}$ (the control parameter of the experiments), the values for the effective fibre bending rigidity and the level of bulk compression (the control parameters of the effective single component model). To achieve this goal, we need to match the output from the experiments ($G_0$, $\gamma_o$ and $\gamma_c$) with the output from the simulations ($G_0 d/\mu_1$, $\gamma_o$ and $\gamma_c$) for every experimental measurement.

To allow quantitative mapping, the experimental moduli expressed in units of Pascal had to be converted to the dimensionless quantities used in the simulations. First, the linear elastic moduli of composite networks were normalized with respect to the modulus found at $c_{HA}$=0 mg/mL. Subsequently, the normalized moduli were multiplied by the linear modulus obtained in the effective single component simulations of this work (Figure 4c) with $\tilde{\kappa}_{eff}$ set to the $\tilde{\kappa}$ values found by mapping measured strain-stiffening curves for the atelocollagen and telocollagen networks on simulation results[1] as described in the Methods section ($\tilde{\kappa} = 3 \cdot 10^{-5}$ and $1 \cdot 10^{-4}$, for the atelocollagen and telocollagen, respectively).

The set of matching simulation parameters is determined in the following way: we start with the two-dimensional simulation phase space with $\tilde{\kappa}_{eff}$ on one axis and the level of bulk compression on the other axis. In this phase space we obtained the linear modulus, the onset strain and the critical strain for 1326 simulation points using the effective single component model. These points form a map for every single output parameter as a function of $\tilde{\kappa}_{eff}$ and the bulk compression. For every experimental measurement (i.e. every hyaluronan concentration), we want to find the point in this simulation phase space that best matches all the experimental output parameters. Because the dependence of the simulation output parameters on the control parameters is smooth, we can draw contour lines that separate the region in phase space with simulation output parameters that are higher than those found in experiment from the region where simulation output parameters are lower. Supplementary Figure 20 shows an example of how this is done in case of composites of atelocollagen and 5 mg/mL hyaluronan. For most values of $c_{HA}$, the curves corresponding to the experimental $G_0$ and $\gamma_c$ crossed at a single point in the compression-$\tilde{\kappa}_{eff}$ parameter space. We identified the parameters of the simulation closest to this crossing point and considered these to be the simulation parameters on which the experimental values map. We considered only the linear modulus and the critical strain in the mapping and not the onset strain, because the error in obtaining this value is big compared to the relevant range for mapping.

This procedure was performed for composite networks of hyaluronan with telocollagen and with atelocollagen (Supplementary Figure 20). We observed significantly higher compression values for atelocollagen compared to telocollagen, consistent with the predictions from the double network model (Figure 3d). In contrast, the values for $\tilde{\kappa}_{eff}$ are comparable above a hyaluronan concentration of 3 mg/mL, indicating that for both types of collagen the linear modulus is dominated by the hyaluronan matrix above this concentration. Below 3 mg/mL, the scaling of the linear modulus with the hyaluronan concentration is absent or very weak, indicating that here the linear modulus is dominated by the collagen fibre network.

## Supplementary Figures

Supplementary Figure 1: Hyaluronan does not alter collagen network architecture.

Supplementary Figure 2: Particle tracking measurements reveal that the hyaluronan matrix is uniform.

Supplementary Figure 3: Hyaluronan does not associate with collagen.

Supplementary Figure 4: Hyaluronan forms crosslinked networks at concentrations of 3 mg/mL and above.

Supplementary Figure 5: Effect of delayed hyaluronan crosslinking on the mechanics of the composite.

Supplementary Figure 6: Polymerization curves and normal stress evolution for pure collagen, pure hyaluronan, and composite system.

Supplementary Figure 7: Ablation experiments show how the surrounding stressed hyaluronan matrix affects the recoil of collagen fibres.

Supplementary Figure 8: Two-component simulations: network initialization.

Supplementary Figure 9: Choosing two-component simulation parameters to mimic experimental conditions.

Supplementary Figure 10: Simulated two-component stiffening response when varying independently matrix stiffness and matrix prestress.

Supplementary Figure 11: Mechanical properties of simulated two-component networks with $\tilde{\kappa}=3\cdot10^{-5}$, corresponding to atelocollagen.

Supplementary Figure 12: Linear mechanical enhancement in the double network simulations.

Supplementary Figure 13: Changing the fibre rigidity by including or removing the telopeptide regions of collagen.

Supplementary Figure 14: Effect of hyaluronan on telocollagen network mechanics.

Supplementary Figure 15: Power law relationship between onset and critical strain for composite systems with varying hyaluronan concentration.

Supplementary Figure 16: Mechanical properties of simulated two-component networks with $\tilde{\kappa}=1\cdot10^{-4}$, corresponding to telocollagen.

Supplementary Figure 17: Rescaling stress-stiffening curves onto a master curve.

Supplementary Figure 18: Establishing a mechanical balance in effective single-component network simulations by resetting the angles.

Supplementary Figure 19: Mechanical response of an effective single-component model.

Supplementary Figure 20: Mapping effective single-component network simulations on the experimental system.

Supplementary Figure 21 | Experimental determination of onset and critical strain.

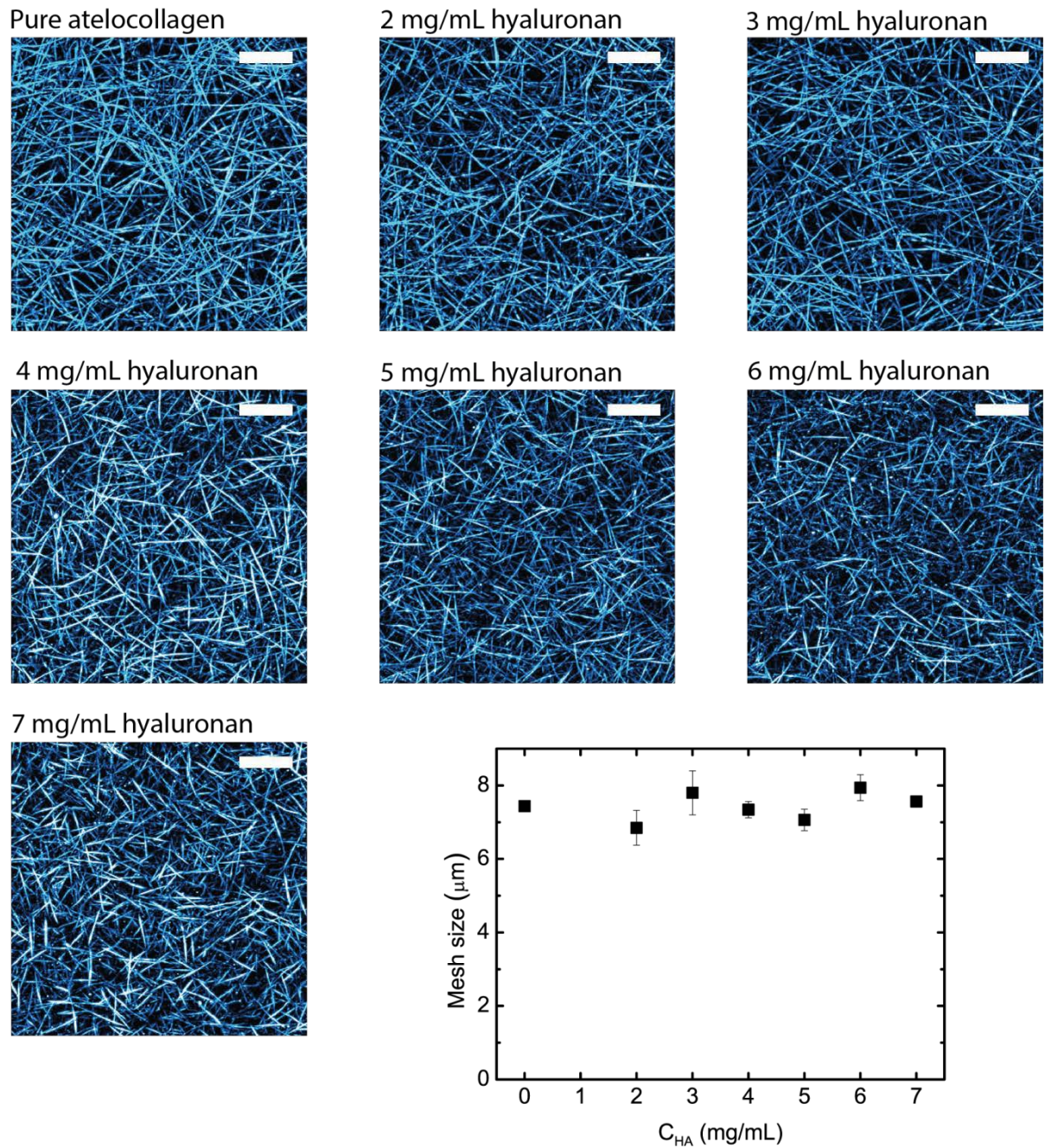

**Supplementary Figure 1 | Hyaluronan does not alter collagen network architecture.** Confocal reflectance images of collagen in the presence of hyaluronan at concentrations varying from 0 to 7 mg/mL, together with a quantitative analysis of the average mesh size, reveal no substantial architectural changes in the collagen architecture. Scale bar indicates 10 μm.

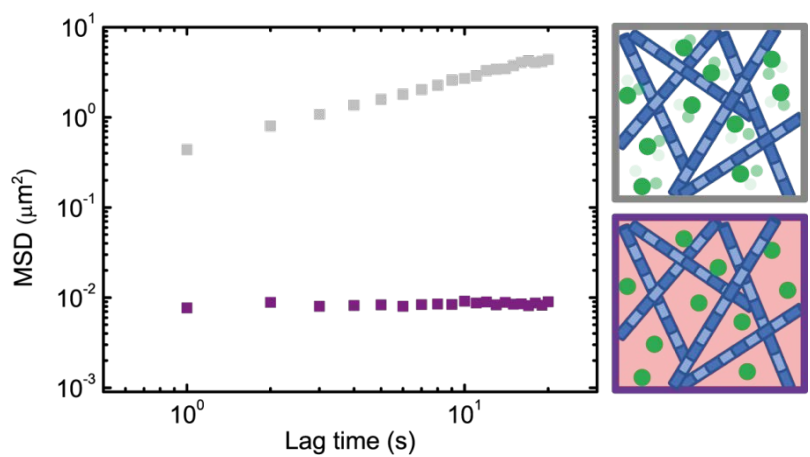

**Supplementary Figure 2 | Particle tracking measurements reveal that the hyaluronan matrix is uniform.** Mean square displacement (MSD) of particles (shown in green in the schematics) embedded in pure collagen networks versus a composite system (1 mg/mL collagen and 4 mg/mL hyaluronan). The particles freely move in the solvent phase between the collagen pores in pure collagen networks (grey symbols), whereas they are immobilized in the hyaluronan mesh in the composite system (purple symbols).

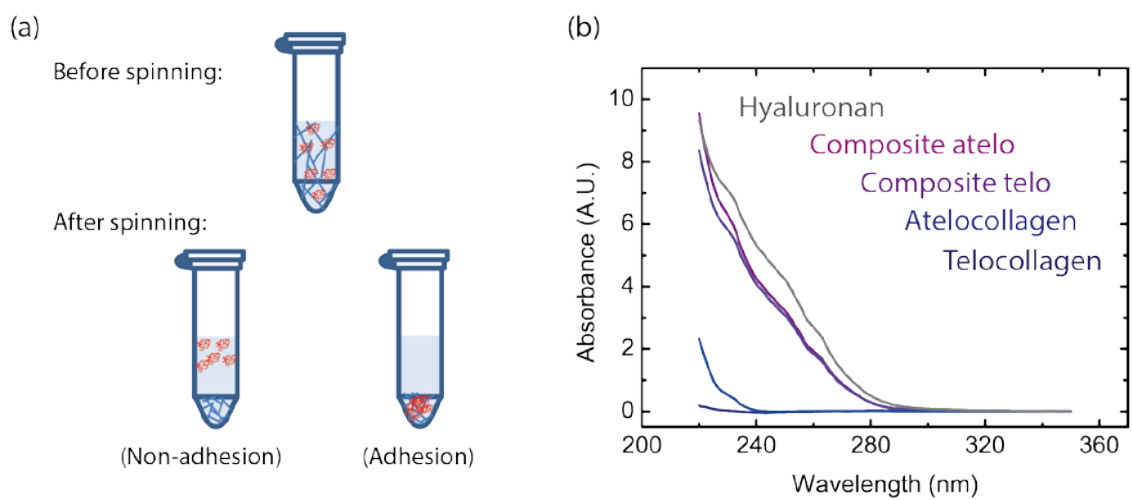

**Supplementary Figure 3 | Hyaluronan does not associate with collagen**. (a) Schematic of the co-sedimentation assay, showing the composite network before centrifugation (upper panel) and after centrifugation (lower panel). Hyaluronan remains in the supernatant in case it does not bind collagen, and ends up in the pellet together with collagen in case it does bind. (b) Light absorbance as a function of wavelength for the supernatant obtained after centrifugation of a composite network with either atelocollagen or telocollagen (1 mg/mL collagen, 4 mg/mL hyaluronan), compared to the absorbance of pure solutions of hyaluronan (4 mg/mL), atelocollagen (1 mg/mL), and telocollagen (1 mg/mL). Since the absorbance spectrum of the supernatant resembles that of hyaluronan, it can be concluded that hyaluronan does not exhibit appreciable binding to collagen.

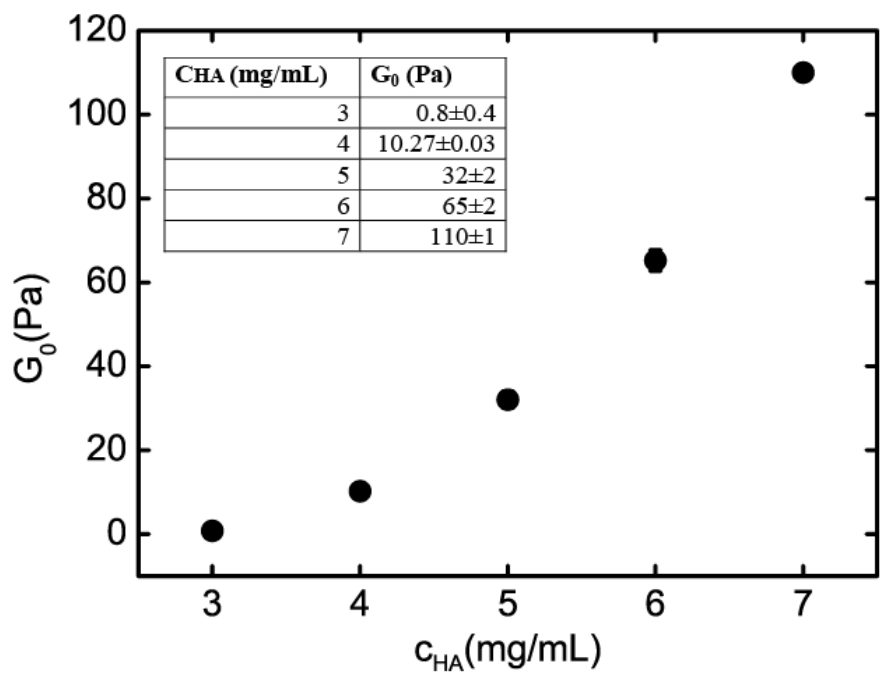

| CHA (mg/mL) | $G_0$ (Pa) |
|---|---|
| 3 | 0.8±0.4 |
| 4 | 10.27±0.03 |
| 5 | 32±2 |
| 6 | 65±2 |
| 7 | 110±1 |

**Supplementary Figure 4 | Hyaluronan forms crosslinked networks at concentrations of 3 mg/mL and above.** Linear elastic modulus (and standard error of the mean based on 3 repeats) of crosslinked hyaluronan networks as a function of hyaluronan concentration, over the range examined in the main text. At concentrations below 3 mg/mL, we were not able to robustly measure the linear modulus, due to sensitivity limitations of the instrument. The molar ratio between the PEGDA crosslinker and the hyaluronan polymer is kept constant.

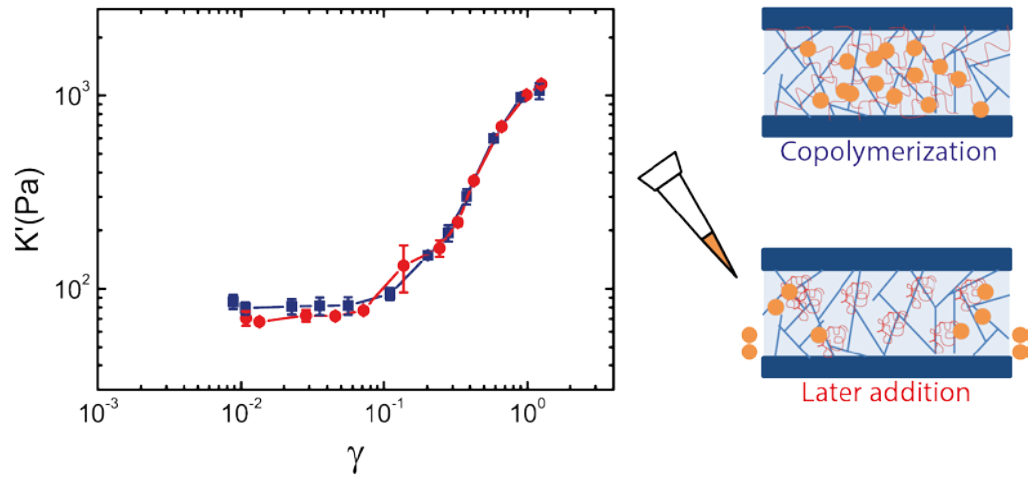

**Supplementary Figure 5 | Effect of delayed hyaluronan crosslinking on the mechanics of the composite.** Strain-stiffening curves of a composite network (1 mg/mL collagen and 5 mg/mL hyaluronan) where the PEGDA crosslinker is included before collagen polymerization (blue symbols, see schematic on the right, where the crosslinkers are indicated in orange) or after collagen polymerization (red) are nearly identical. The average values on the basis of three independent repeats are $G_0 = 90 \pm 7$ Pa, $\gamma_0 = 0.22\pm0.02$, and $\gamma_c = 0.43\pm0.10$ for co-gelation, compared to $G_0 = 80 \pm 10$ Pa, $\gamma_0 = 0.2\pm0.01$, and $\gamma_c = 0.44\pm0.12$ for later crosslinker addition. Note that the moduli are slightly different from the data shown in the main text due to the use of a smaller cone plate geometry (20 mm instead of 40 mm) and a longer polymerization time (6 hours instead of 2).

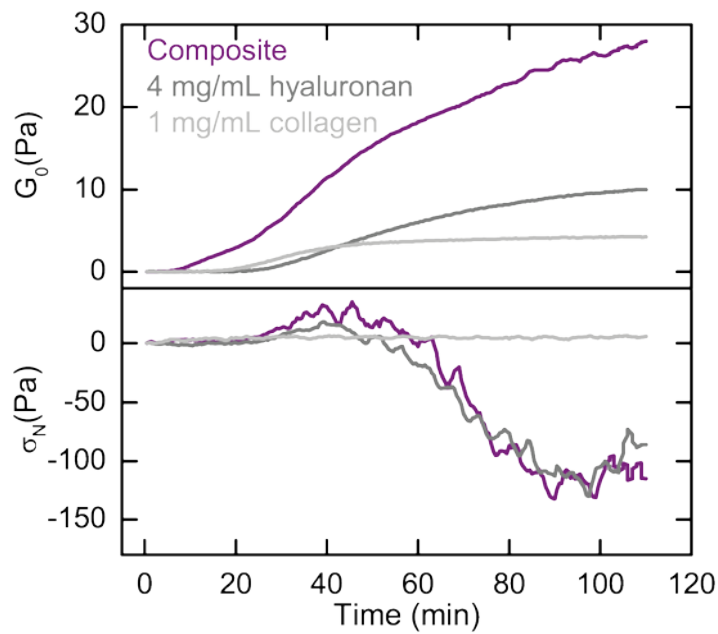

**Supplementary Figure 6 | Polymerization curves and normal stress evolution for pure collagen, pure hyaluronan, and composite system.** Top panel: Time evolution of the linear elastic shear modulus during polymerization for a composite system composed of 1 mg/mL collagen and 4 mg/mL hyaluronan (purple, top-most curve), pure collagen at 1 mg/mL (light gray), and pure hyaluronan 4 mg/mL (dark grey). Bottom panel: Corresponding time evolution of the normal stress as a function of polymerization time. Note that no measurable normal stress is observed in the pure collagen system. By contrast, the normal stress for the pure hyaluronan network and the collagen-hyaluronan composite first increases and then decreases, finally reaching a negative value indicative of contractile prestress.

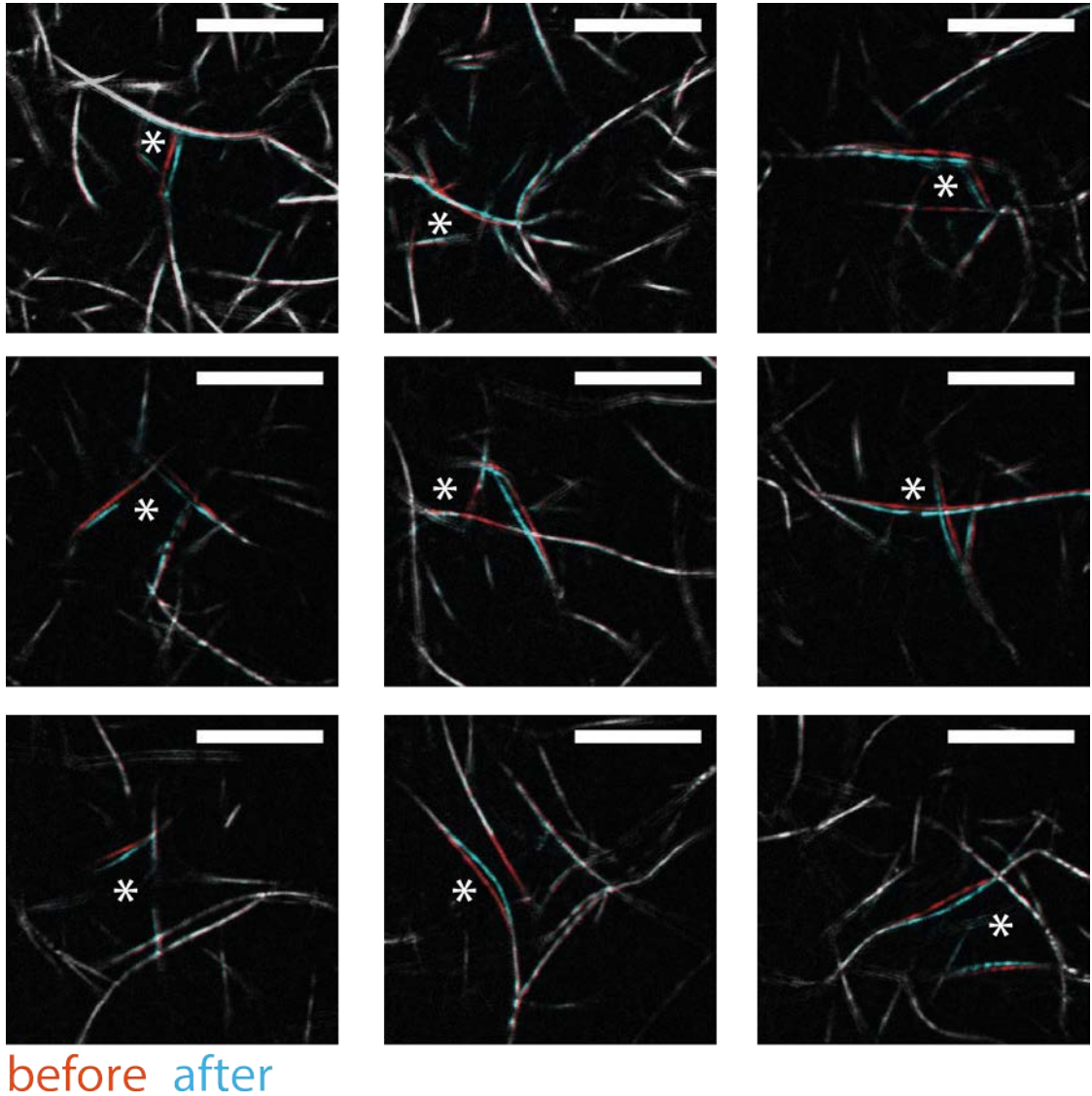

before after

**Supplementary Figure 7 | Ablation experiments show how the surrounding stressed hyaluronan matrix affects the recoil of collagen fibres.** Time projections of composite networks, showing only the collagen network imaged with label-free confocal reflectance microscopy, before and after localized ablation of the background hyaluronan network. The networks are composed of 1 mg/mL collagen and 4 mg/mL hyaluronan. Scale bars indicate 10 μm and asterisks represent the approximate location of the ablation spot.

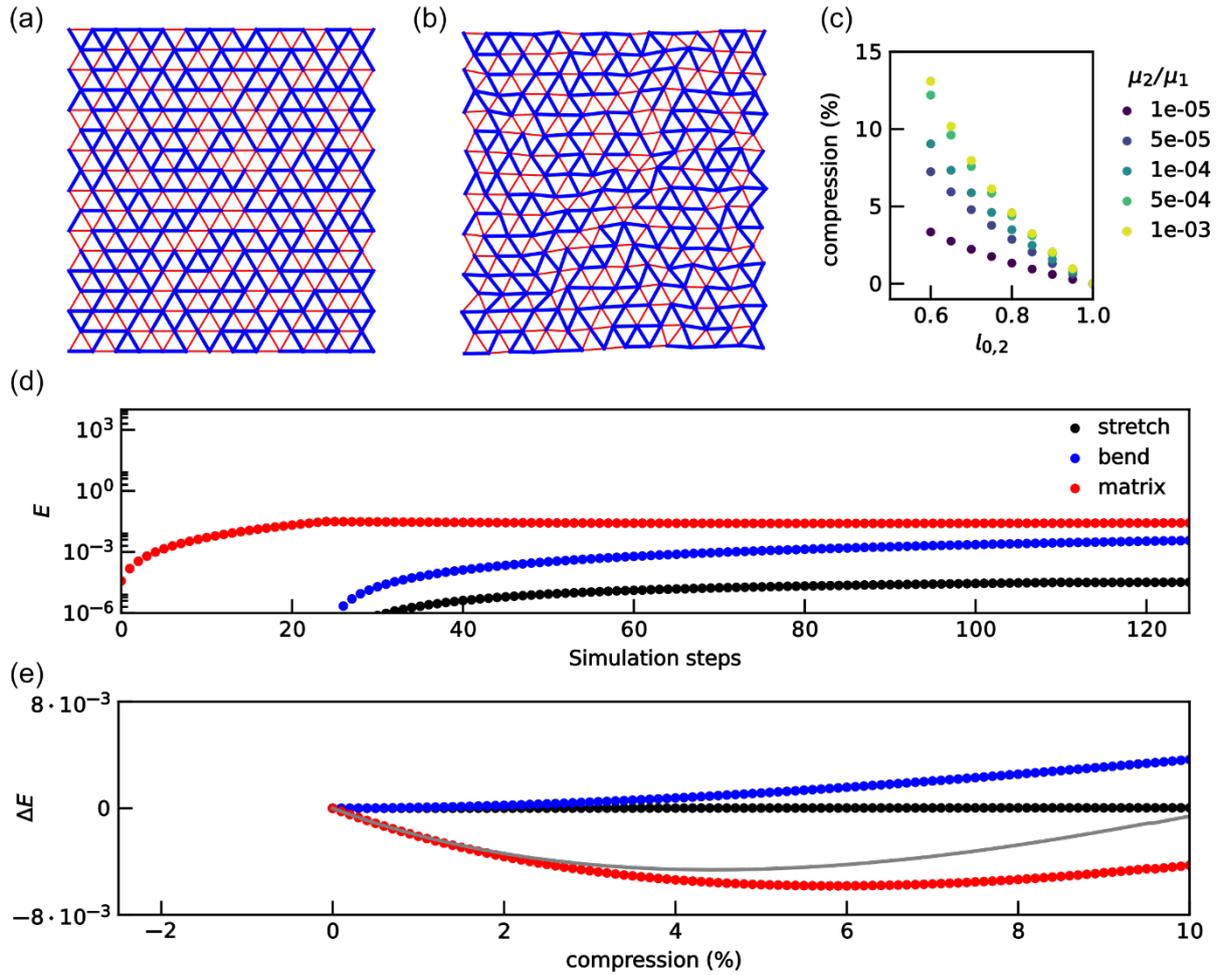

**Supplementary Figure 8 | Two-component simulations: network initialization.** Snapshots of a double network before (a) and after (b) bulk compression by 3.4% bulk strain. (c) Equilibrium values of bulk compression as a function of $l_{0,2}$ for $\tilde{\kappa} = 3 \cdot 10^{-5}$, shown for different values of $\mu_2/\mu_1$. (d) Total energy of the composite system as a function of the number of simulation steps during compression, decomposed in the stretching and bending energy of the collagen network and the stretching energy associated with the hyaluronan matrix. (e) Energy change upon bulk compression with respect to the reference state (0% bulk compression, $l_{0,2}$=0.75). The total energy is indicated with a grey line. The equilibrium state corresponds to the state of minimum energy. The panel shows an example obtained for a diluted triangular network with ⟨z⟩=3.4, L=50, $\tilde{\kappa} = 3 \cdot 10^{-5}$, $\tilde{\mu}_2 = 3 \cdot 10^{-5}$ and $l_{0,2} = 0.75d$, where mechanical equilibrium occurs at a compression of 4.4%.

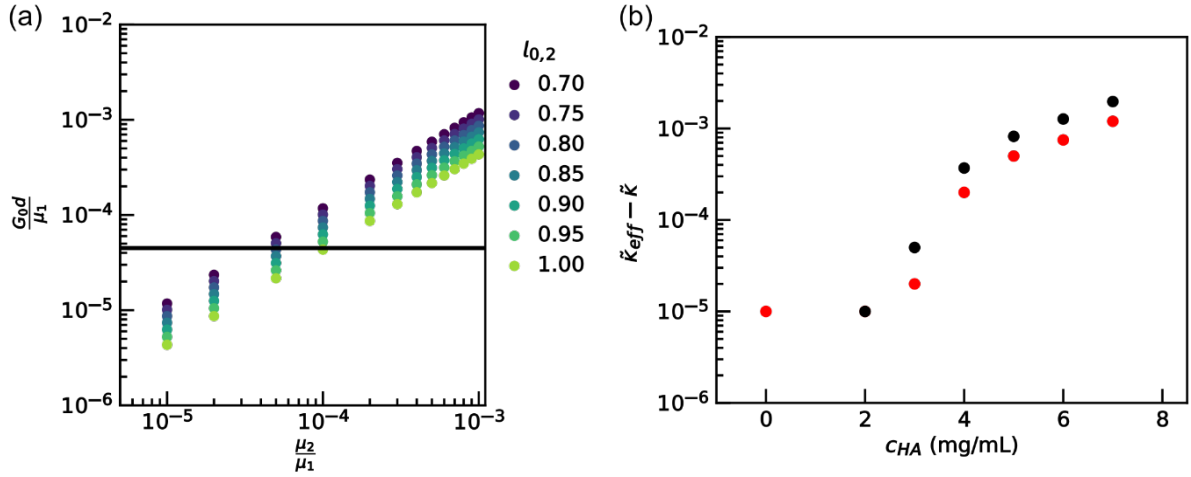

**Supplementary Figure 9 | Choosing two-component simulation parameters to mimic experimental conditions.** (a) Linear shear modulus for homogeneous triangular networks of springs as a function of $\mu_2/\mu_1$. The black line indicates the linear modulus of a diluted triangular lattice with average connectivity 3.4 and $\tilde{\kappa} = 3 \cdot 10^{-5}$. (b) In the effective single component model, the effective bending rigidity[11] follows the equation $\tilde{\kappa}_{eff} = \tilde{\kappa} + \tilde{\mu}_2 d^2$, hence $\tilde{\kappa}_{eff} - \tilde{\kappa}$ should be equal to $\tilde{\mu}_2$ when $d$ is equal to unity. In (b) the difference between the effective bending rigidity found in the mapping of the results for the composite networks and the bending constant for bare atelocollagen (red, $\tilde{\kappa} = 3 \cdot 10^{-5}$) and telocollagen (black, $\tilde{\kappa} = 1 \cdot 10^{-4}$) is shown. The matrix stiffness found in this way ranges from $10^{-5}$ to $10^{-3}$, which corresponds with the range of $\tilde{\mu}_2$ values explored in the double network simulations.

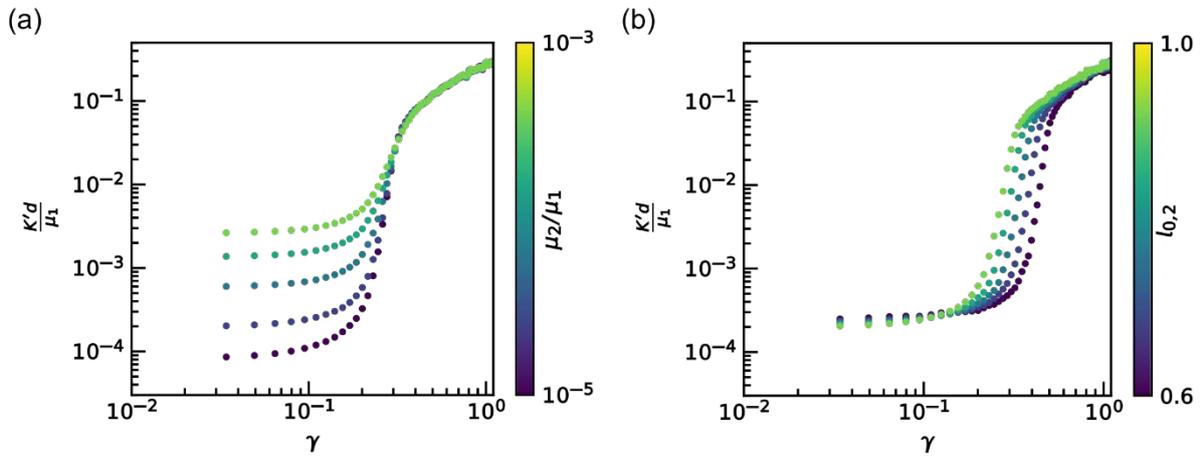

**Supplementary Figure 10 | Simulated two-component stiffening response when varying independently matrix stiffness and matrix prestress.** Modulus versus strain curves for (a) varying matrix stiffness at fixed $l_{0,2} = 0.75d$ and (b) varying matrix rest length at fixed $\mu_2/\mu_1 = 10^{-4}$. Results obtained using diluted triangular networks with $\langle z \rangle = 3.4$, L=50, $N_{config}=20$, $\tilde{\kappa} = 3 \cdot 10^{-5}$, and $\tilde{\mu}_2 = 3 \cdot 10^{-5}$.

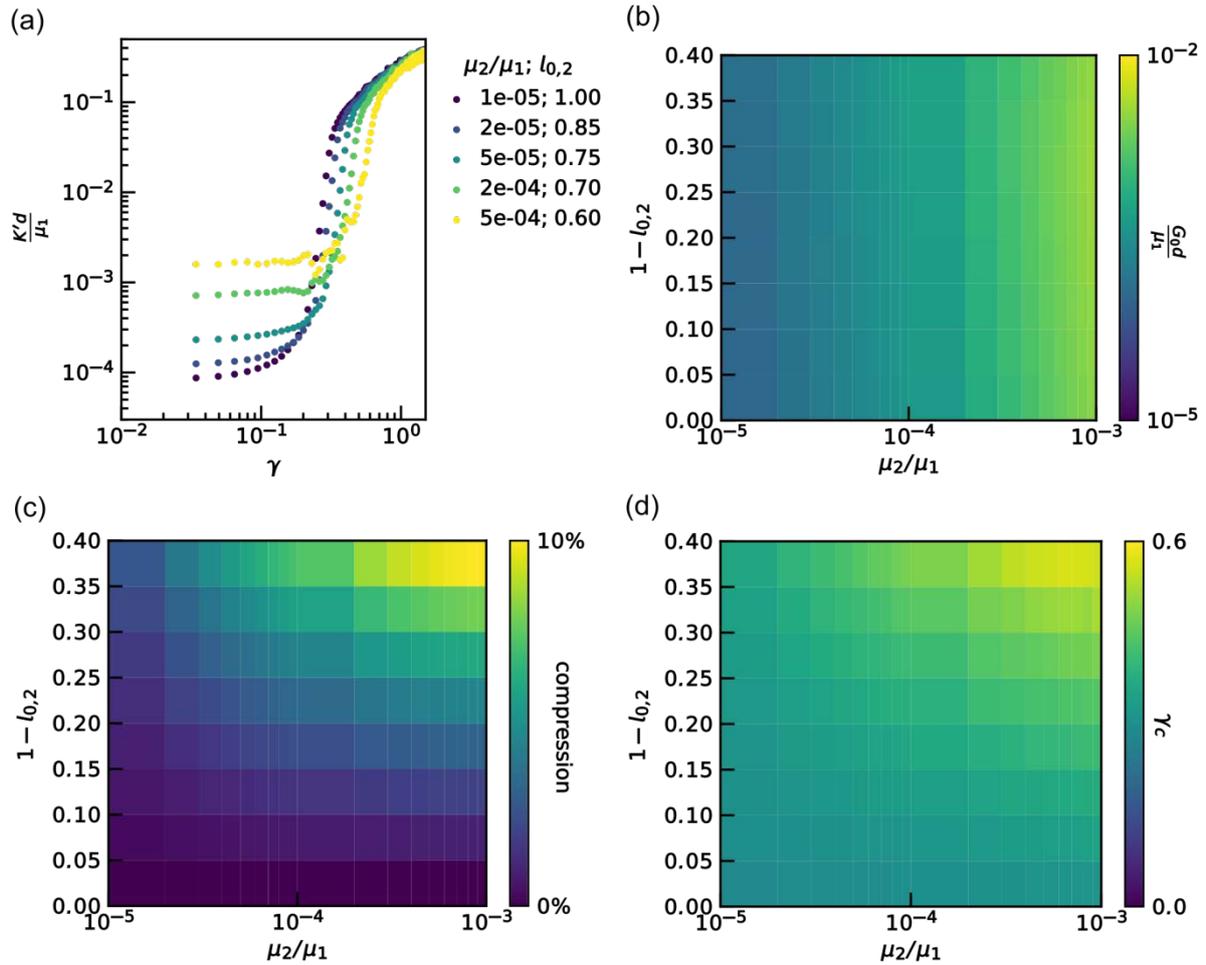

**Supplementary Figure 11 | Mechanical properties of simulated two-component networks with $\tilde{\kappa} = 3 \cdot 10^{-5}$, corresponding to atelocollagen.** (a) Example of modulus versus strain curves obtained by changing simultaneously the stretching constant and the rest length of the hyaluronan matrix. (b-d) Phase space representations of the (b) linear elastic shear modulus, (c) bulk compressive strain at mechanical equilibrium, and (d) critical strain, shown as a function of matrix stiffness relative to fibre network stiffness and of the reduction of matrix segment rest length.

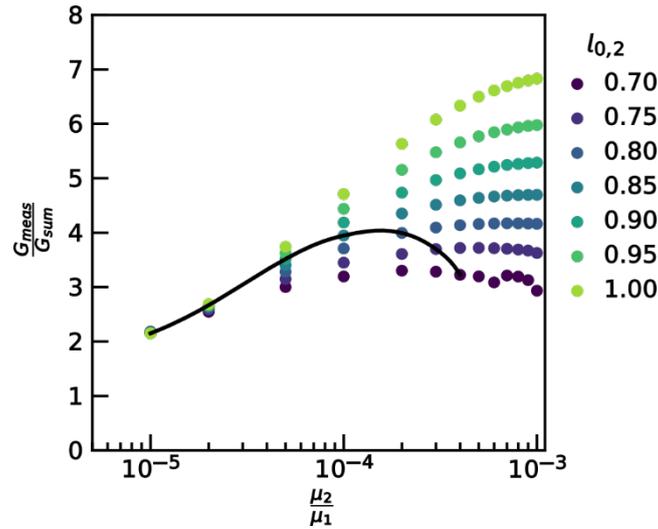

**Supplementary Figure 12 | Linear mechanical enhancement in the double network simulations.** The linear mechanical enhancement is calculated for the double network simulations as $G_{meas}/G_{sum}$, with $G_{meas}$ the modulus obtained in the double network simulations, and $G_{sum}$ obtained as the sum of the modulus of the fibre network ($\langle z \rangle = 3.4$, $\tilde{\kappa} = 3 \cdot 10^{-5}$) and that of the matrix (homogeneous triangular lattice with stiffness $\tilde{\mu}_2$). Without internal stress ($l_{0,2} = 1.00$), the linear mechanical enhancement is proportional to the matrix stiffness. For small $\tilde{\mu}_2$, the linear mechanical enhancement depends weakly on the matrix rest length $l_{0,2}$. However, for large $\tilde{\mu}_2$ there is a strong dependence on $l_{0,2}$. Experimentally, a peak is observed in the linear mechanical enhancement as a function of $c_{HA}$, as shown in Figure 1c. Qualitatively, we observe a similar trend in the simulation data. We start with the notion that the mechanical response as a function of hyaluronan concentration matches qualitatively with a simultaneous increase of matrix stiffness and decrease of rest length in the experiment (see figure 3b). From Supplementary Figure 9b we can estimate that the stiffness of the matrix ranges from $10^{-5}$ to $10^{-3}$. In addition, the quantitative mapping predicts a compression of 10% at a hyaluronan concentration of 5 mg/mL. In the double network simulations, this corresponds to $l_0$ values between 1.00 and 0.70, as can be concluded from Supplementary Figure 8c. The black line shows that, if the rest length decreases with increasing stiffness roughly following these ranges, a peak in the mechanical enhancement is present.

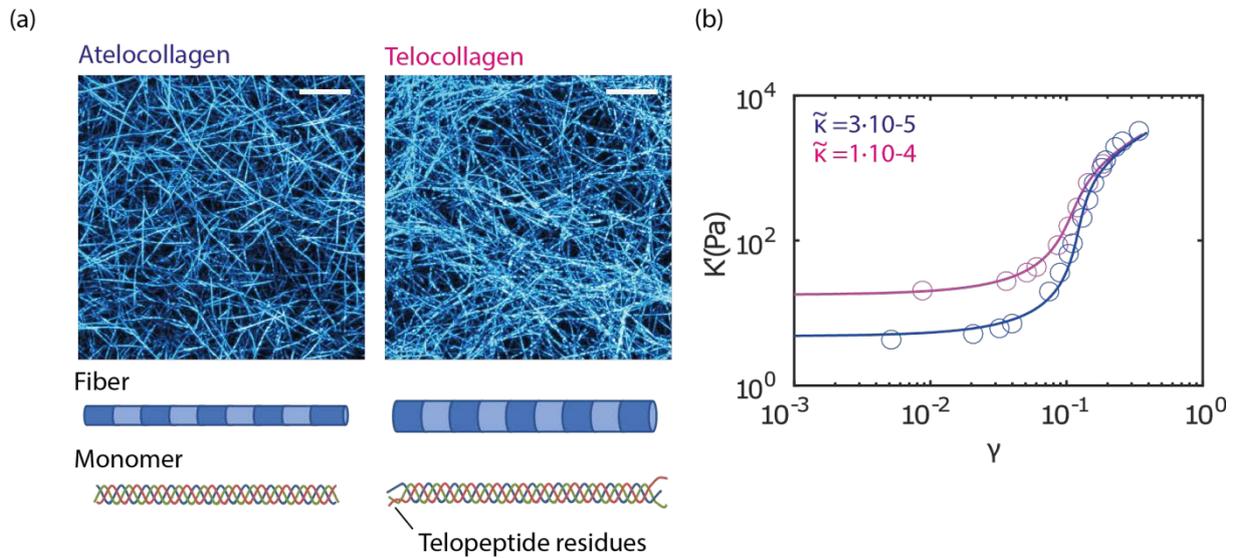

**Supplementary Figure 13 | Changing the fibre rigidity by including or removing the telopeptide regions of collagen.** (a) Z-stack projections of atelo and telocollagen networks (1 mg/mL) imaged with confocal reflectance microscopy, together with schematics showing the difference between the two at the fibre level (different fibre diameter) and molecular level (absence or presence of telopeptides that mediate crosslinking between adjacent collagen monomers). (b) Fits (lines) of the athermal fibrous network model[7] to the measured strain-stiffening curves of the two types of networks (circles) reveal a different dimensionless fibre rigidity for collagen and atelocollagen (inset).

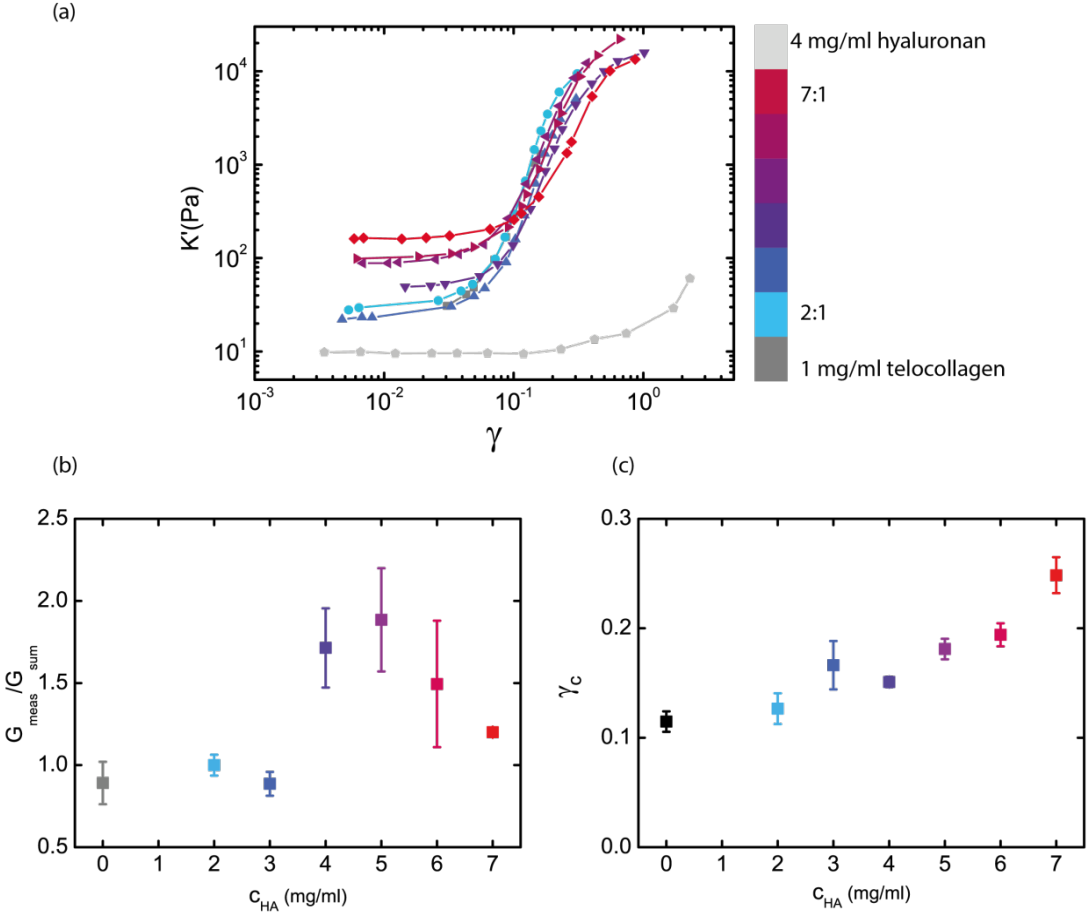

**Supplementary Figure 14 | Effect of hyaluronan on telocollagen network mechanics.** (a) Strain stiffening curves of a pure telocollagen network (1 mg/mL) and composites with increasing hyaluronan concentration (see colour bar on the right). (b) Linear mechanical enhancement and (c) critical strain measured as a function of hyaluronan concentration. Error bars in (b) represent the error obtained with error propagation, while for (c) they are obtained by considering the standard error on the mean of at least three independent measurements.

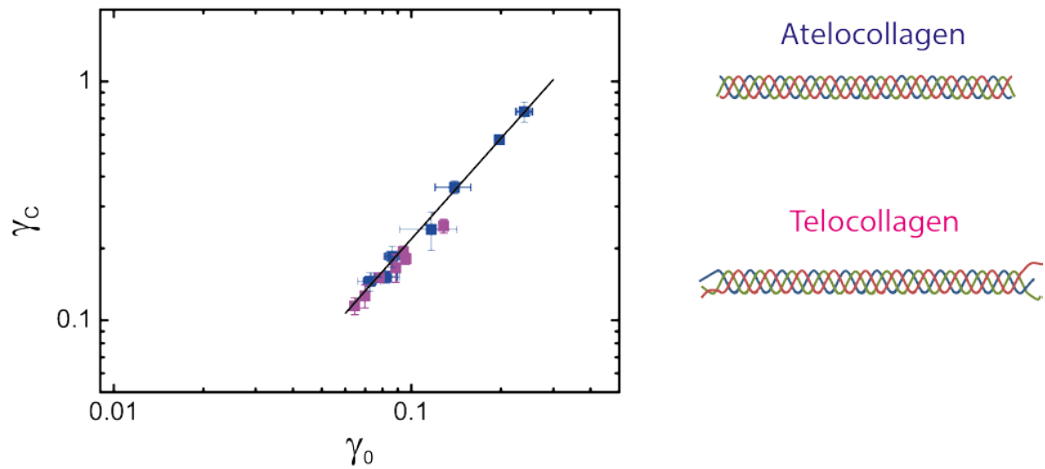

**Supplementary Figure 15 | Power law relationship between onset and critical strain for composite systems with varying hyaluronan concentration.** The relationship between onset and critical strain follows a power law, consistent with prior observations for pure collagen systems[1]. Solid black line indicates a power-law with exponent 1.4.

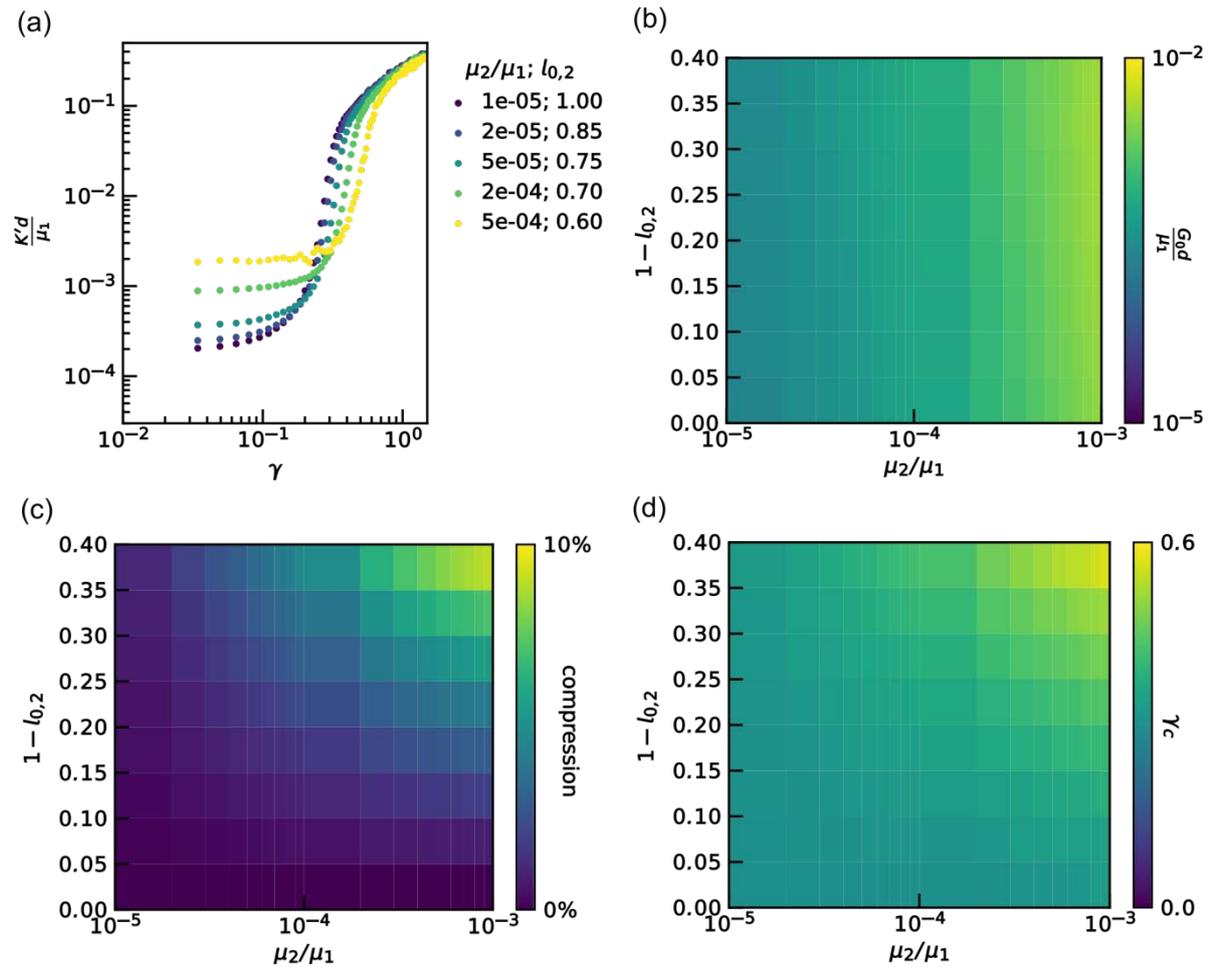

**Supplementary Figure 16 | Mechanical properties of simulated composite networks with $\tilde{\kappa} = 1 \cdot 10^{-4}$, corresponding to telocollagen.** (a) Example strain-stiffening curves. (b) Linear modulus, (c) compression at mechanical equilibrium, (d) critical strain, shown in terms of the two matrix-related control parameters: matrix stiffness $\mu_2$ and rest length reduction $1-l_{0,2}$.

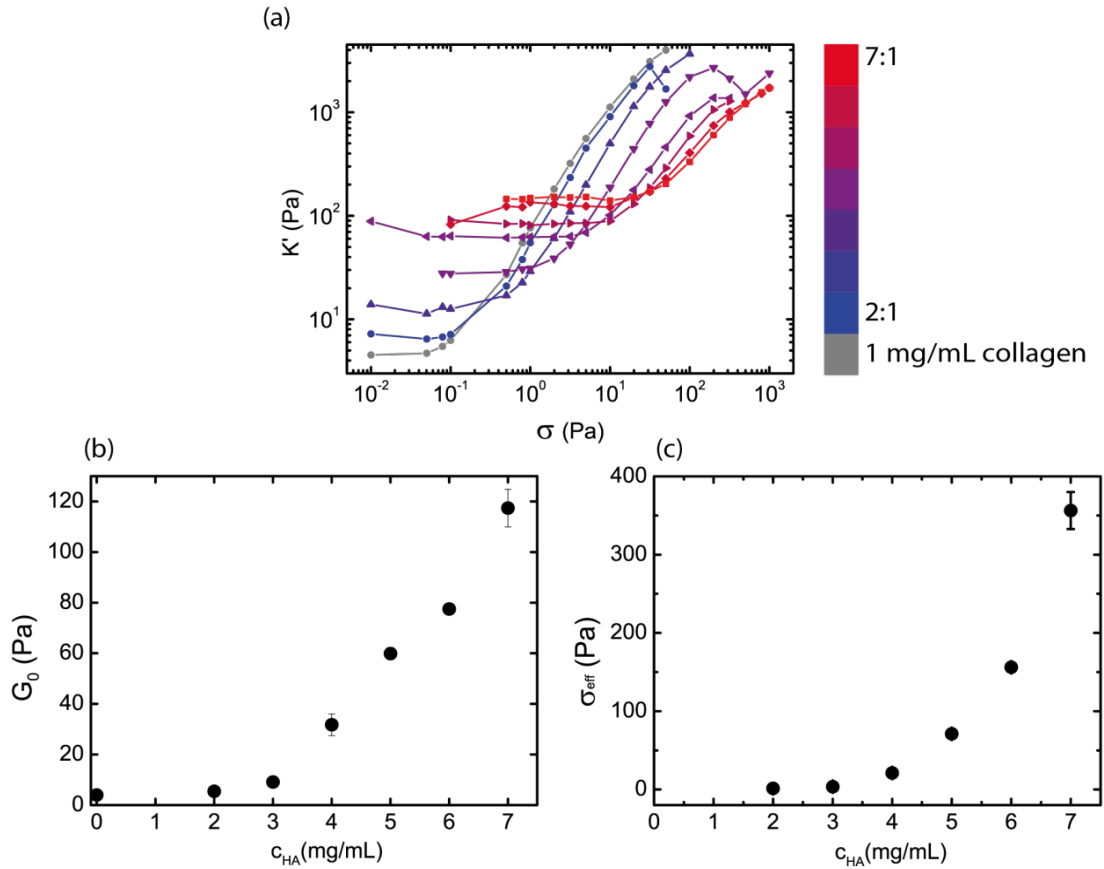

**Supplementary Figure 17 | Rescaling stress-stiffening curves onto a master curve.** (a) Stress-stiffening curves of composite networks composed of 1 mg/ml atelocollagen and various concentrations of hyaluronan, which were used for obtaining the master curve in Figure 4. (b) Values of $G_0$ used to rescale curves on the y-axis, and (c) values of effective stress $\sigma_{eff}$ used to rescale curves on the x-axis, both shown as a function of hyaluronan concentration. Error bars represent standard error on the mean of at least three independent measurements.

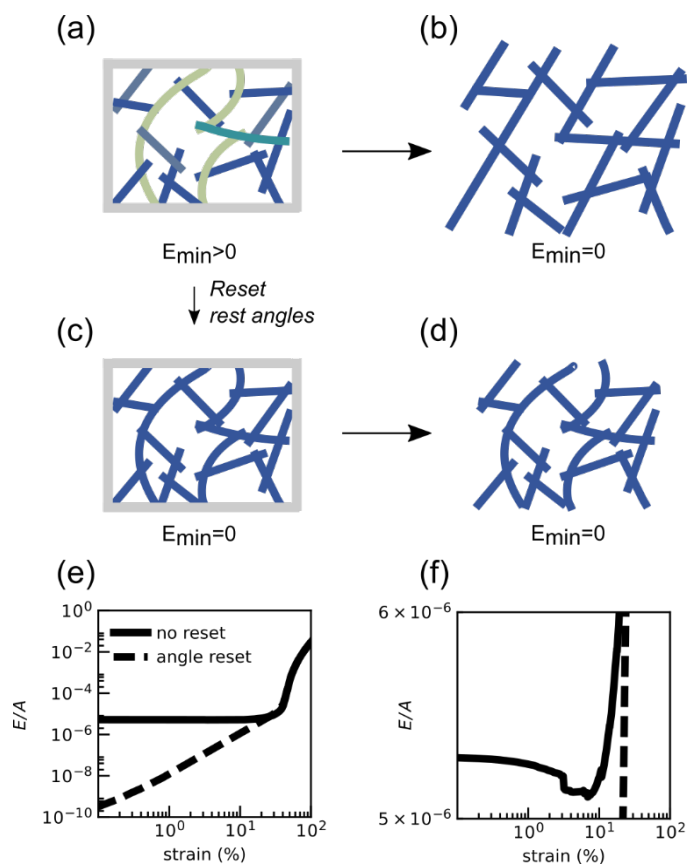

**Supplementary Figure 18 | Establishing a mechanical balance in effective single-component network simulations by resetting the fibre angles.** (a) If a network is compressed by forcing it to a smaller system size, the fibres will bend. This will increase the bending energy of some fibres, as indicated by the colour coding from blue (low energy) to yellow (high energy). (b) If the boundaries are removed, the system will go back to its original state in order to alleviate the excess bending energy. This is in contrast with the double network simulations, where a mechanical balance is established between the fibres and the matrix. (c) To prevent the system from moving back to its original state, we can reset the rest angles of the fibres in the compressed state. This also lowers the energy of the system, because now the fibres are in their rest state in this particular configuration. (d) Now, if the boundaries are removed, the system will retain its shape. (e) The necessity of the angle reset becomes clear if we compare the development of the total energy of the system, divided by the area as a function of shear ($<z>=3.2$, $L=30$, $N_{config}=5$). If the angles are not reset after compression, the energy starts at a high value, after which it goes down. If the angles are reset before shear, the behaviour is markedly different: the energy starts at zero and increases linearly up to the onset of strain stiffening. At the onset of strain stiffening, the curves with and without an angle reset converge, because here the stretching terms start to dominate. (f) Zoom-in of panel (e) showing that the total energy decreases upon increasing strain.

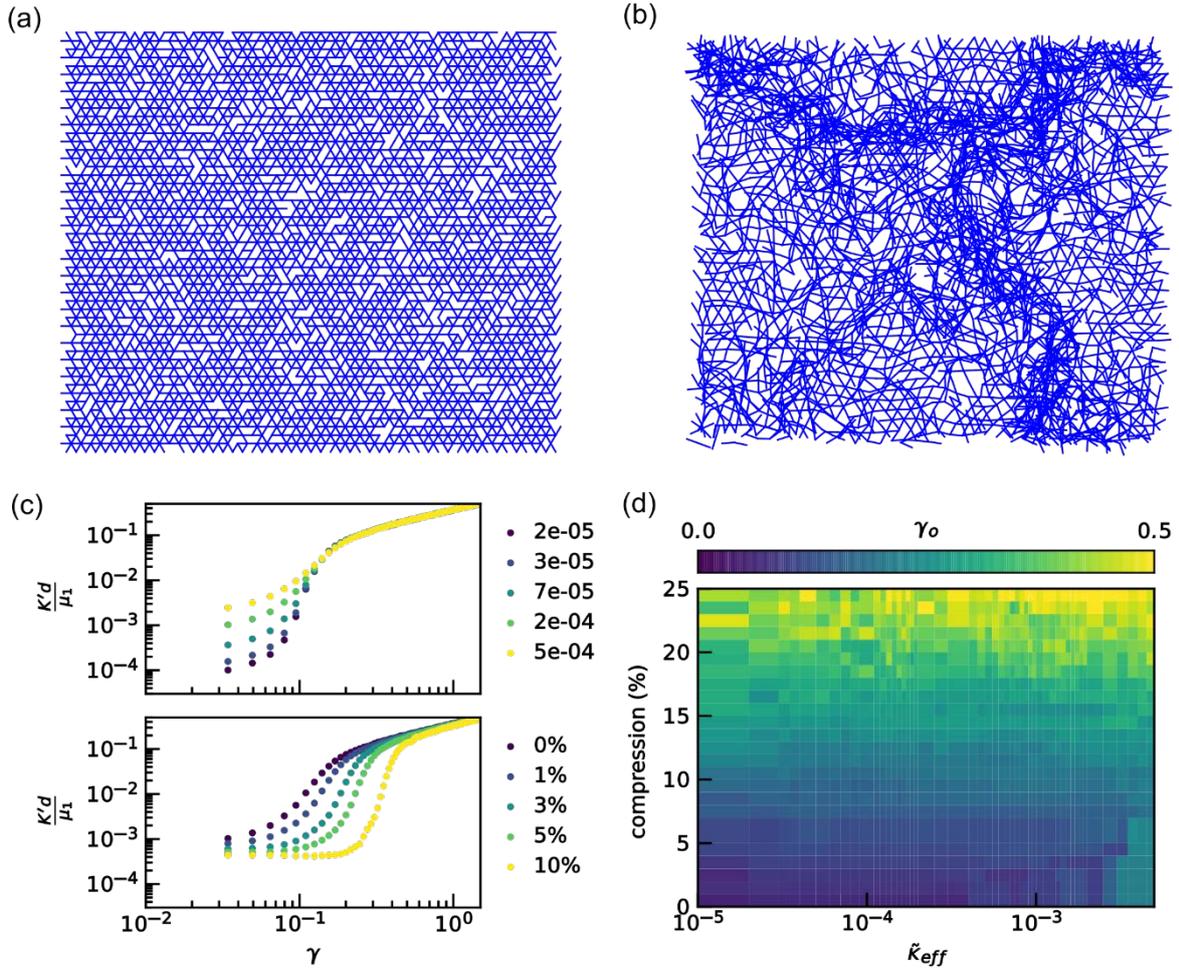

**Supplementary Figure 19 | Mechanical response of an effective single-component model.** Snapshots of a phantom network with $\langle z \rangle$=3.2 and $\tilde{\kappa}_{\text{eff}} = 1 \cdot 10^{-4}$ (L=50) (a) before compression and (b) after 25% isotropic compression. (c) Modulus versus strain for different effective fibre rigidity $\tilde{\kappa}_{\text{eff}}$ (indicated in the legend) without compression (top panel), and modulus for fixed $\tilde{\kappa}_{\text{eff}}$=1·$10^{-4}$ with different levels of compression (as indicated in the legend, bottom panel). (d) Map of the onset strain as a function of the bulk compressive strain and $\tilde{\kappa}_{\text{eff}}$ ($N_{\text{config}}$=5).

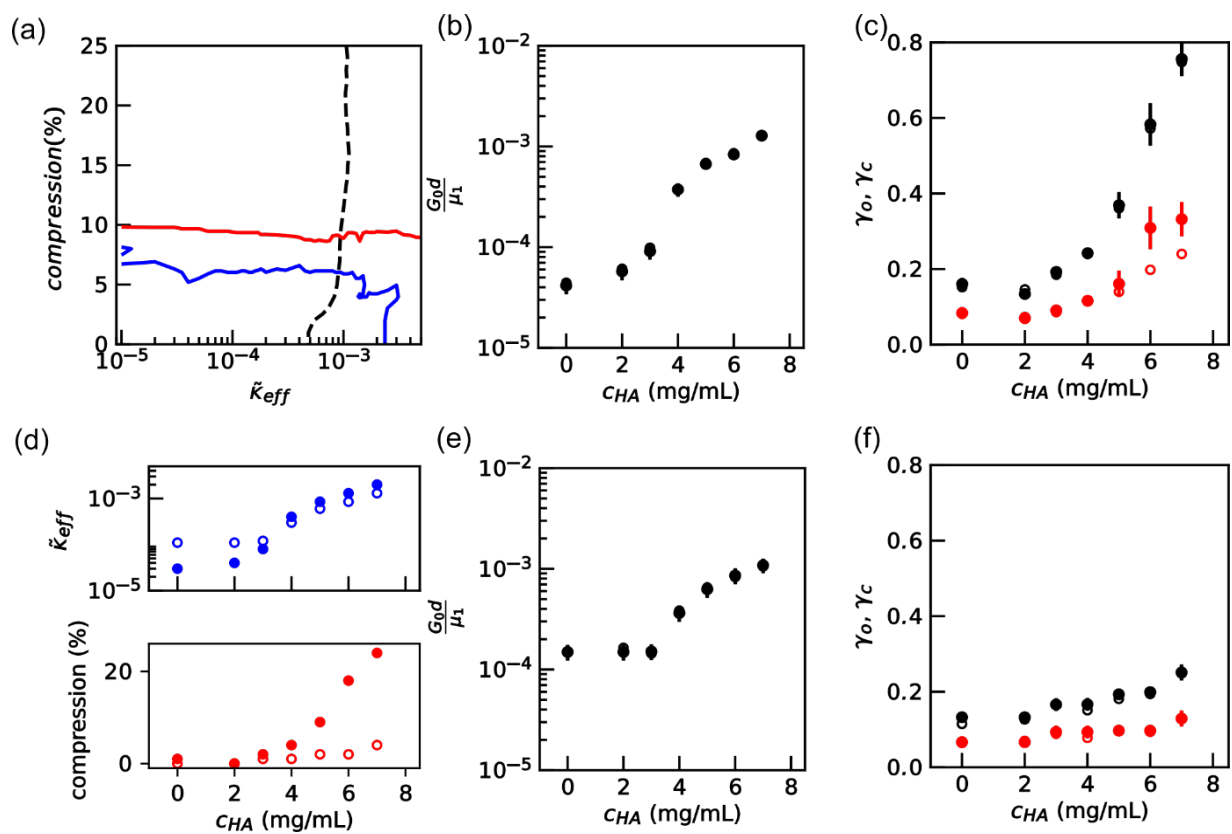

**Supplementary Figure 20 | Mapping effective single-component network simulations on the experimental system.** (a) Example of the mapping procedure. Using the simulation maps for the linear modulus, onset and critical strain (like Supplementary Figure 19) we draw contour lines corresponding to the experimental values (1 mg/mL atelocollagen, 5 mg/mL hyaluronan) in the compression - effective bending rigidity plane. The dashed black line corresponds to the linear modulus, red line to the critical strain and blue line to onset strain. The point where the lines corresponding to the linear modulus and the critical strain cross identifies the simulation parameters that correspond to the experimental composite networks with given hyaluronan concentration. The onset strain is noisier than the linear modulus and the critical strain and is therefore not considered for the mapping. (b,c) Mapping of the linear modulus, onset strain (red) and critical strain (black) measured for atelocollagen composites on the simulation results. Solid symbols correspond to experimental values and open symbols to simulation values (often the two are overlapping). (d) Experimental results mapped on the simulation parameters of the effective single-component model for atelocollagen (solid symbols) and telocollagen (open symbols) as explained in SI section 1.4. (e,f) Same as panels (b,c) for telocollagen. To allow for mapping, the linear modulus found in the experiments was rescaled as described in the Supplementary Information.

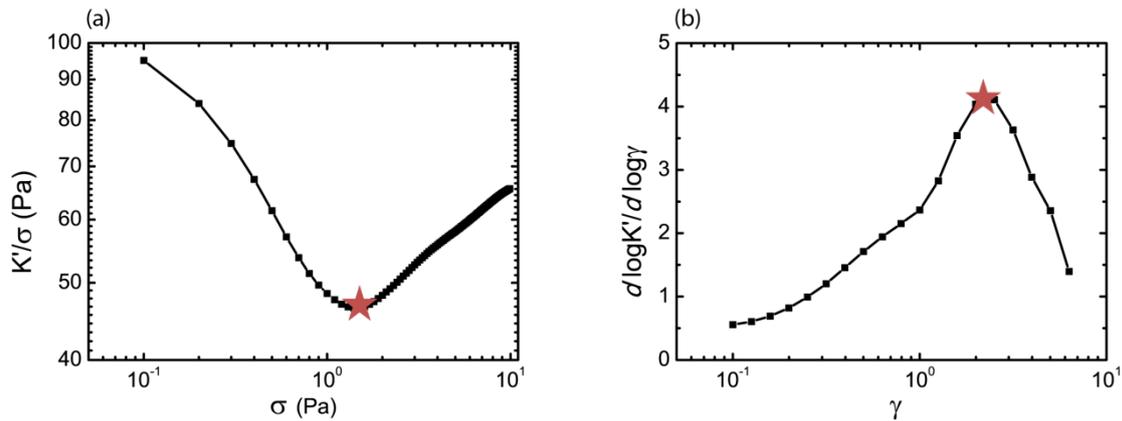

**Supplementary Figure 21 | Experimental determination of onset and critical strain.** Representative curves show how onset and critical strain are determined. Following a previously introduced definition[7], the onset strain is computed by taking the stress value at which the curve K'/σ as a function of σ attains its minimum and then considering the corresponding strain (a). The critical strain corresponds to the strain value at which the derivative of log K with respect to log γ attains its maximum value (b).

## REFERENCES

1) Broedersz, C.P., and MacKintosh, F.C. Modeling semiflexible polymer networks. *Rev. Mod. Phys.* **86,** 995 (2014).

2) Sharma, A. *et al.* Strain-controlled criticality governs the nonlinear mechanics of fibre networks. *Nat. Phys.* **12,** 584–587 (2016).

3) Feng, J., Levine, H., Mao, X., Sander, L.M. Nonlinear Elasticity of Disordered Fibre Networks. *Soft Matter* **12,** 1419-1424 (2016).

4) Rens, R., Vahabi, M., Licup, A.J., Mackintosh, F.C., Sharma, A. Nonlinear Mechanics of Athermal Branched Biopolymer Networks. *J. Phys. Chem. B* **26,** 5831–5841 (2016).

5) Sharma. A., *et al.* Strain-driven criticality underlies nonlinear mechanics of fibrous networks Phys. Rev. E 94, 042407

6) Picu, R.C., Mechanics of Random Fiber Networks – a Review. *Soft Matter* **7,** 6768-6785 (2011).

7) Jansen, K.A., *et al.* The Role of Network Architecture in Collagen Mechanics. *Biophys. J.* **114,** 2665-2678 (2018).

8) Lindström, S.B., Vader, D.A., Kulachenko, A., Weitz, D.A. Biopolymer network geometries: Characterization, regeneration and elastic properties. *Phys. Rev. E* **82,** 051905 (2010).

9) Broedersz, C.P., Mao, X., Lubensky, T.C., MacKintosh, F.C. Criticality and isostaticity in fibre networks. *Nature Phys.* **7,** 983–988 (2011).

10) Bitzek, E., Koskinen, P., Gähler, F., Moseler, M., Gumbsch, P. Structural Relaxation Made Simple. *Phys. Rev. Lett.* **97,** 170201 (2006).

11) van Doorn, J.M., Lageschaar, L., Sprakel, J., van der Gucht, J. Criticality and mechanical enhancement in composite fiber networks. *Phys. Rev. E* **95,** 042503 (2017).

12) van Oosten, A.S.S.G. *et al.* Uncoupling shear and uniaxial elastic moduli of semiflexible biopolymer networks: compression-softening and stretch-stiffening. *Sci. Rep.* **6,** 19270 (2016).

13) Licup, A.J., Sharma, A., MacKintosh, F.C. Elastic regimes of subisostatic athermal fiber networks. *Phys. Rev. E* **93,** 012407 (2016).